\documentclass[12pt]{article}
\usepackage{scicite}
\usepackage{times}

\usepackage{amssymb,amsmath}
\usepackage[USenglish]{babel}
\usepackage{graphicx}
\usepackage{multirow}
\usepackage{amsfonts}
\usepackage{textcomp}
\usepackage[dvipsnames]{xcolor}
\usepackage[utf8]{inputenc}
\usepackage[version=4]{mhchem}
\usepackage{makecell}
\usepackage{float}
\usepackage{calrsfs}
\usepackage{siunitx}
\let\svqty\qty
\usepackage{physics}
\let\qty\svqty
\usepackage{lineno}

\definecolor{darkgoldenrod}{rgb}{0.72, 0.53, 0.04}

\AtBeginDocument{\RenewCommandCopy\qty\SI}
\ExplSyntaxOn
\msg_redirect_name:nnn { siunitx } { physics-pkg } { none }
\ExplSyntaxOff

\newcommand{\kB}{k_{\text{B}}}
\newcommand{\muB}{\mu_{\text{B}}}
\newcommand{\expe}{\mathrm{e}}
\newcommand{\cplxi}{\mathrm{i}}

\newenvironment{sciabstract}{%
\begin{quote} \bf}
{\end{quote}}

\title{Terahertz Emission from Diamond Nitrogen-Vacancy Centers} 

\author 
{S\'{a}ndor~Kollarics,$^{1,2}$ Bence~G\'{a}bor~M\'{a}rkus,$^{3,2}$ Robin~Kucsera,$^{1}$ Gerg\H{o}~Thiering,$^{2}$ \\ \'{A}d\'{a}m~Gali,$^{2,4}$ Gergely~N\'{e}meth,$^{2}$ Katalin~Kamar\'{a}s,$^{2,5}$L\'{a}szl\'{o}~Forr\'{o},$^{3,6}$ Ferenc~Simon$^{1,2\ast}$\\
\\
\normalsize{$^{1}$Department of Physics, Institute of Physics, and ELKH-BME Condensed Matter Research Group,}\\
\normalsize{Budapest University of Technology and Economics, M\H{u}egyetem rkp. 3., H-1111 Budapest, Hungary,}\\
\normalsize{$^{2}$Institute for Solid State Physics and Optics,}\\
\normalsize{HUN-REN Wigner Research Centre for Physics, PO. Box 49, H-1525, Hungary}\\
\normalsize{$^{3}$Stavropoulos Center for Complex Quantum Matter, Department of Physics and Astronomy,}\\
\normalsize{University of Notre Dame, Notre Dame, Indiana 46556, USA}\\
\normalsize{$^{4}$Department of Atomic Physics, Institute of Physics,}\\
\normalsize{Budapest University of Technology and Economics, M\H{u}egyetem rkp. 3., H-1111 Budapest, Hungary}\\
\normalsize{$^{5}$Institute of Technical Physics and Materials Science,}\\
\normalsize{Centre for Energy Research, P.O. Box 49, H-1525 Budapest, Hungary}\\
\normalsize{$^{6}$Laboratory of Physics of Complex Matter,}\\
\normalsize{\'{E}cole Polytechnique F\'{e}d\'{e}rale de Lausanne, Lausanne CH-1015, Switzerland}\\
\\
\normalsize{$^\ast$To whom correspondence should be addressed; E-mail:  simon.ferenc@ttk.bme.hu}
}

\date{}

\begin{document} 


\baselineskip24pt


\maketitle


\begin{sciabstract}
Coherent light sources emitting in the terahertz range are highly sought after for fundamental research and applications. THz lasers rely on achieving population inversion. We demonstrate the generation of THz radiation using nitrogen-vacancy (NV) centers in a diamond single crystal. Population inversion is achieved through the Zeeman splitting of the $S = 1$ state in $15$ T, resulting in a splitting of {\color{black} $0.42\ \text{THz}$}, where the middle $S_z=0$ sublevel is selectively pumped by visible light. To detect the THz radiation, we utilize a phase-sensitive THz setup, optimized for electron spin resonance measurements (ESR). We determine the spin-lattice relaxation time up to $15$ T using the light-induced ESR measurement, which shows the dominance of phonon-mediated relaxation and the high efficacy of the population inversion. The THz radiation is tunable by the magnetic field, thus these findings may lead to the next generation of tunable coherent THz sources.
\end{sciabstract}


\section*{Introduction}

Despite considerable efforts, the domain known as the THz gap ($0.1-10~\text{THz}$) remains one of the less explored and exploited ranges of the electromagnetic spectrum. This is mainly due to a lack of coherent radiation sources between electronics-based microwave devices (operating range below {\color{black} $0.1\ \text{THz}$}) and infrared lasers, operating above $100~\text{cm}^{-1}$ ($3~\text{THz}$). The quest for coherent terahertz sources is motivated by both fundamental research and foreseen applications. Examples of the earlier include the study of correlated matter (superconductors \cite{SupraTHz}, density wave systems \cite{CDW_THz}, exotic magnets \cite{THz_magnons_ACSPhot2018}), non-equilibrium charge dynamics \cite{TimeResTHz}, and otherwise metastable states \cite{MetastableTHz}. For applications, THz radiation is expected to revolutionize various areas including broad-band communication, medical imaging, fusion research, and security applications \cite{THzRoadmap}.

{\color{black}Incoherent} THz sources are black-body radiators such as mercury lamps or light-gated Auston switches \cite{AustonSwitch}, coherent sources include laser frequency mixing- \cite{THzFreqMixing} and Josephson-effect based devices \cite{SuperconductorTHzSource}, resonant-tunneling diodes \cite{ResTunDiode}, electron tubes \cite{BWO_THz}, quantum-cascade lasers \cite{QuantumCascadeLasers}, and relativistic-electron sources using synchrotrons \cite{THz_sources_RepProgPhys2006} or free-electron lasers \cite{THzFEL}. Much as these devices satisfy various requirements (including high power, pulsed nature, tunability, etc.), it would be highly desired to present a source that operates based on the same basic laser principle that was originally used in ruby \cite{MaimanLaser}: a solid-state system, optically pumped to population inversion, where efficient stimulated emission occurs.

Of the known solid-state systems with suitable energy level structure, the nitrogen-vacancy center in diamond (more precisely its negative charge state NV(-) with $S=1$ \cite{Gruber1997}) is unique: it can be efficiently pumped with visible light to selectively populate the $S_z=0$ level of the ground state triplet while depleting the $S_z=\pm 1$ levels \cite{Manson2006NVtheory}. In addition, the spin-lattice relaxation time is about a few ms at room temperature \cite{Nivotsev2005,Pham2011,Jarmola2012}, which is long enough to maintain the light-induced population. Thus in {\color{black}a sufficiently strong} magnetic field, population inversion occurs between the upper-lying $S_z=0$ and the lower-lying $S_z=-1$ state. In addition, the NV center is known to be extremely photostable: no photobleaching occurs, which is usually the limiting factor for {\color{black}single-molecule emitters, e.g., the rhodamine dyes}. These properties were exploited to yield the first NV center-based maser \cite{Breeze2018}, which operates at $9.2~\text{GHz}$ and an NV based amplifier at $18$ GHz \cite{NV_Maser_18GHz}. We note that non-diamond based alternative quantum emitter systems are also actively researched \cite{Dyakonov_SiC_maser}.

Herein, we observe emission at $0.42~\text{THz}$ (and at $0.21$~THz) from NV centers in single crystals of diamond subjected to a $15~\text{T}$ ($7.5$ T) magnetic field. A phase-sensitive THz detector attests that the emitted THz photons are coherent with the incoming excitation, whose energy matches the Zeeman splitting between the two states with the inverted population. We also study the sensitive dependence of the population inversion efficiency of the NV center axis orientation with respect to the external magnetic field. We show that the spin-selective intersystem crossing (ISC) rates and thus the optical spin-polarization processes are primarily determined by the orbitals within the $C_{3v}$ crystal field symmetry even at {\color{black}even at magnetic fields up to $15\ \text{T}$}. The spin-lattice relaxation time of the NV centers down to $20~\text{K}$ is measured using a double-modulated electron spin resonance experiment. The result excludes the presence of a relaxation time speeding-up due to magnetic field, thus population inversion can be effectively maintained, which makes the NV system a promising candidate for a coherent THz source.

\section*{Results}

\begin{figure}[!ht]
	\centering
	\includegraphics*[width=1.00\linewidth]{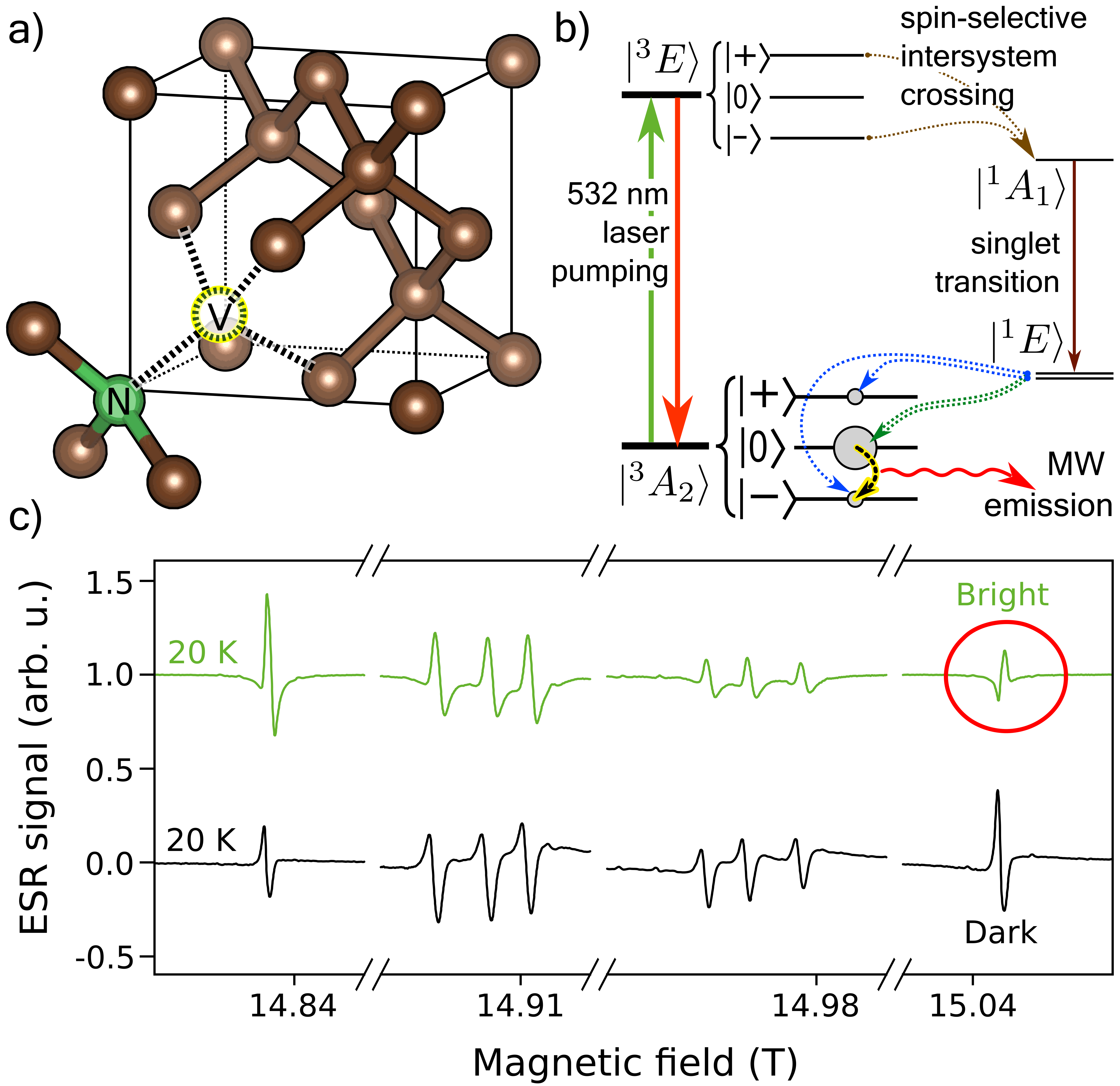}
	\caption{{\color{black}\textbf{Structure, energy level diagram and electron spin resonance spectra of NV centers.}} a) Structure of the nitrogen-vacancy center in diamond. b) Energy level diagram and population of the three sublevels of the $S=1$ ground state triplet in the dark and under intensive illumination (bright). We show the non-radiative intersystem crossing (ISC) rates from the excited triplet $\ket{^3 E}$ into the metastable singlet state $\ket{s}$ and towards the ground state triplet $\ket{^3 A_2}$. The ISC from the $\ket{\pm}$ sublevels of $\ket{^3 E}$ is much stronger than from the $\ket{0}$. This, under intensive laser irradiation, leads to the pumping of all electrons to the $\ket{0}$ state of $\ket{^3A_2}$. This is also confirmed by our experiments. In a sufficiently high magnetic field, this corresponds to a population inverted scenario. c) High-field ESR spectra of the NV centers at $20~\text{K}$ in the dark and under laser illumination. The ESR signal of charge neutral nitrogen atoms (also known as the P1 signal) was removed from the center of the spectrum to improve clarity. Note that the uppermost signal (indicated by a circle) reverses, which is clear evidence of a population inversion.}
\label{Fig1}
\end{figure}

Nitrogen is a common substitutional center in diamond. In Type Ib diamond, nitrogen atoms are dispersed and they can form the so-called NV center if a vacancy is captured on a neighboring site. Its structure is shown in Fig. 1a. The vacancies are naturally present in the samples or are induced by irradiation and these become mobile at high temperatures, thus efficiently forming the thermally stable NV center \cite{Davies1976NVanneling, Loubser1978, Mita1996}. The interest is focused on the negatively charged state of the center or NV(-), where the source of the electron is an ionized nitrogen atom \cite{Larsson2008NVchargetransfer}. For the sake of simplicity, we denote the NV(-) centers as NV in the following.

The ground state of the NV center is an $S=1$ triplet state, where in the absence of an external magnetic field the $S_z=0$ level is separated from the degenerate $S_z=\pm 1$ levels by the zero-field splitting (ZFS) parameter of $D/h\approx~2.87~\text{GHz}$. These levels are split in a strong magnetic field. The Jablonski diagram in Fig. 1b shows the excited and intermediate levels, as well as the possible transitions. The NV centers absorb light between $450~\text{and } 637~\text{nm}$. Although in dark conditions the respective populations of the ground state triplet are given by a Boltzmann distribution, under the intensive irradiation, the $S_z=0$ level is selectively populated {\color{black}due to the spin-orbit coupling induced selective ISC \cite{Thiering_2017}}, whereas the $S_z=\pm 1$ levels are depleted \cite{Harrison2004spinpolarization,Manson2006NVtheory}. This causes a population inversion between the $S_z=-1$ and $S_z=0$ levels in magnetic fields above $\sim 100~\text{mT}$, which leads to a stimulated microwave emission (or masing effect) when the frequency of the energy splitting lies in the microwave range ($B\lesssim 1~\text{T}$). This effect opens the possibility for a tunable maser source \cite{Breeze2018, NV_Maser_18GHz} and for a THz maser or THz amplifier in a high magnetic field. A rigorous theoretical description can be found in the SI.

{\color{black}Figure} 1c shows the high-field/high-frequency electron spin resonance (ESR) spectrum of the NV centers using THz radiation of {\color{black} $0.42\ \text{THz}$}. The instrument setup and the sample are discussed in the Methods section. In brief, the home-built high-field ESR instrument has a sensitivity of {\color{black}$3\cdot10^{10}\ \text{spins/}10^{-4}\text{T}\sqrt{\text{Hz}}$} and it uses a phase-coherent homodyne detection \cite{Nafradi2008a,Nafradi2008b}. The sample contains {\color{black}$12\ \text{ppm}$} of NV centers, which is one of the highest concentrations reported to date \cite{Kollarics2022}. 

The ESR experiment was performed in the dark and under laser illumination, where a $532~\text{nm}$ laser at $100~\text{mW}$ power and with a beam diameter of about $3~\text{mm}$ excited a sample of similar area. Conventional ESR \cite{SlichterBook} is based on observing the resonant absorption of coherent radiation between adjacent levels ($\Delta m=1$), which are the eigenstates of the Zeeman Hamiltonian. In a typical ESR experiment, the microwave irradiation frequency is kept constant and the magnetic field is swept as shown in Fig. 1c. It is important to note that the eight observed ESR transitions look to be similar for the dark spectrum, i.e., they show an identical phase. 

The spin Hamiltonian of the NV center reads:
\begin{equation}\label{eq:hamiltonian}
		\hat{\mathcal{H}}=D\left(S_z^2-\frac{1}{3}S\left(S+1\right)\right)+g_\text{e} \muB \boldsymbol{B} \boldsymbol{S}.
\end{equation}

This Hamiltonian is given in a coordinate system whose $z$-axis is fixed to the NV-axis. The first term is the zero-field Hamiltonian with $D/h\approx 2.87~\text{GHz}$. The second term describes the Zeeman effect with the $g$-factor, $g_\text{e}\approx 2.0029$ \cite{Felton2009}, and $\boldsymbol{B}$ can have an arbitrary direction with respect to the NV-axis. The NV-axis can take four inequivalent directions for an arbitrary magnetic field alignment, thus giving rise to eight possible ESR transitions. The ESR spectrum in Fig. 1c corresponds to a magnetic field almost along one of the four NV-axes, thus we observe two transitions well separated from the other six, which lie closer to each other. Our single-crystal diamond has a $\langle111\rangle$ face and the magnetic field lies along this direction.

The striking observation in Fig. 1c is that the signal at the highest magnetic field (indicated by a circle in the Figure) \emph{changes} its sign under the bright conditions. The ESR signal shows derivative Lorentzian curves due to technical reasons \cite{PooleBook}, therefore the sign change means whether 
the Lorentzian starts with an upwards or downwards direction. The same effect was observed previously in a low magnetic field measurement (at around $1~\text{T}$) \cite{Harrison2004spinpolarization} which was also reproduced herein (data shown in the Supplementary Information). ESR spectroscopy is based on the phase coherent detection of microwaves after interaction with the sample \cite{AbragamBook,PooleBook}. In a usual experiment, the microwave radiation is absorbed, therefore the observed ESR signal phase corresponds to a net microwave photon absorption. The sign reversal in an ESR experiment therefore can only correspond to an extra \emph{emission} of microwave photons, which is a consequence of population inversion. We therefore unambiguously observe the population inversion at $15~\text{T}$, corresponding to an emission of {\color{black} $0.42\ \text{THz}$} photons. In turn, this observation opens the way to stimulated emission of THz radiation.

The data in Fig. 1c allow to determine the magnitude of the spin polarization (or that of the population inversion) as the ESR transition signal amplitude is directly proportional to the population difference in the initial and the final state. The observed change in the ESR signal in Fig. 1c is a factor of $2$ for the lowest transition and a factor of $-0.5$ for the highest one under the bright conditions. Were the pumping complete, i.e., the population inversion perfect, we would expect a change of a factor of $6$ for the lowest and $-2.5$ for the highest transition as given in the Supplementary Information. The incomplete saturation of the level populations is the result of a trade-off in our special sample: on one hand, a sample with a high NV concentration is required to be able to perform an ESR experiment at all. On the other hand, it results in a strongly light-absorbing sample, its optical density being $2.7$ at $532~\text{nm}$ (see the Supplementary Information). This means that light-induced pumping is less effective where less optical power is available. This cannot be compensated by increasing the laser power due to sample heating effects.  

\begin{figure*}[!ht]
	\centering
	\includegraphics*[width=0.9\linewidth]{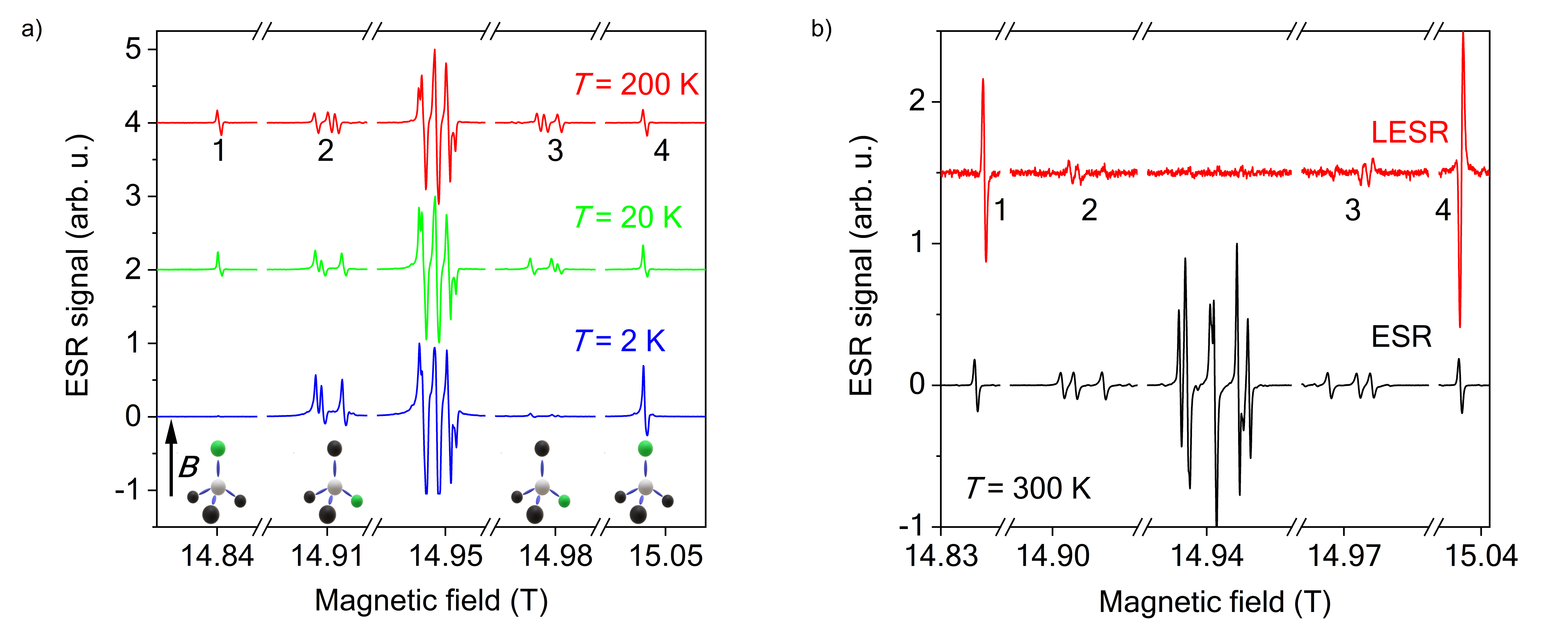} 
	\caption{{\color{black}\textbf{Temperature-dependent electron spin resonance spectra of NV centers and light-induced ESR spectra at room temperature.}} a) Temperature dependence of the conventional ESR signal at $420~\text{GHz}$. The intense lines around $14.95~\text{T}$, i.e., at $g\approx 2$ correspond to the P1 (charge neutral) centers while the other eight lines (denoted as 1-4) belong to negatively charged NV centers. Note that signals 2 and 4 grow on lowering the temperature while 1 and 3 vanish. b) The light-induced ESR (LESR) signal at room temperature and $420~\text{GHz}$ together with the conventional ESR data acquired simultaneously shown as reference. Note that the P1 signal is absent in the LESR spectrum and that resonances 1 and 2 have an \emph{opposite sign} to signals 3 and 4.}
	\label{Fig2}
 \end{figure*}

In order to elucidate the nature of the optimal conditions for a population inversion and the resulting THz emission, we performed temperature-dependent ESR studies as well as light-induced ESR (LESR) studies \cite{ShinarPRL2005}. {\color{black}Figure} 2a shows the temperature dependence of the conventional ESR spectra between $200~\text{and } 2~\text{K}$. We introduce a notation for the observed ESR lines: 1 and 4 denote the transitions for the NV center whose axis is parallel to the external magnetic field, whereas 2 and 3 denote the 3-3 transitions for the other three NV centers whose axes have a high angle with the external magnetic field (close to the tetrahedral angle of $109.5^{\circ}$). The exact resonance energies can be obtained by diagonalizing the spin Hamiltonian in a given magnetic field. However, at high magnetic fields (well above the aforementioned $D/g_{\text{e}}\mu_{\text{B}}\approx 0.1~\text{T}$), when the DC magnetic field is along one of the NV-axes and the other three NV orientations form the tetrahedral angle with it, the resonance frequencies simplify to
\begin{subequations}\label{eq:resfreqs}
    \begin{align}
    	 f_1&= \frac{1}{h} \left(B'+D \right) & \rightarrow B_1&= B_0-D', \\
	   f_2&= \frac{1}{h}\left(B'+\frac{\sqrt{3}}{4}D\right) & \rightarrow B_2&= B_0-\frac{\sqrt{3}}{4}D', \\
	   f_3&= \frac{1}{h}\left(B'-\frac{\sqrt{3}}{4}D\right) & \rightarrow B_3&= B_0+\frac{\sqrt{3}}{4}D', \\
	   f_4&= \frac{1}{h}\left(B'-D\right) & \rightarrow B_4&= B_0+D',
    \end{align}
\end{subequations}
where $f_2$ and $f_3$ are triply degenerate resonances. We introduced the magnetic field in energy units (${B'=g_{\text{e}}\mu_{\text{B}}B}$) and $h$ is Planck's constant. As in continuous-wave ESR the irradiation frequency ($f_0$) is constant and the external magnetic field is swept, the resonance frequencies can be rewritten to resonant fields in the high-field indicated by the arrow.

There we introduced $B_0=\frac{h}{g_{\text{e}}\mu_{\text{B}}}f_0$ and the ZFS parameter in magnetic field units, $D'=D/g_{\text{e}}\mu_{\text{B}}$.

The resonant fields, given in the right side of Eq. (\ref{eq:resfreqs}), are numbered in the same order as in Fig. 2a and they also correspond to the notation in Fig. 3. The ESR spectra presented in Fig. 2a change with decreasing temperature: signals 2 and 4 increase, while signals 1 and 3 vanish. This indicates that signals 2 and 4 correspond to transitions between the lowest and the middle states, whereas signals 1 and 3 correspond to transitions between the middle and highest states.

\begin{figure*}[!ht]
	\centering
	\includegraphics*[width=\linewidth]{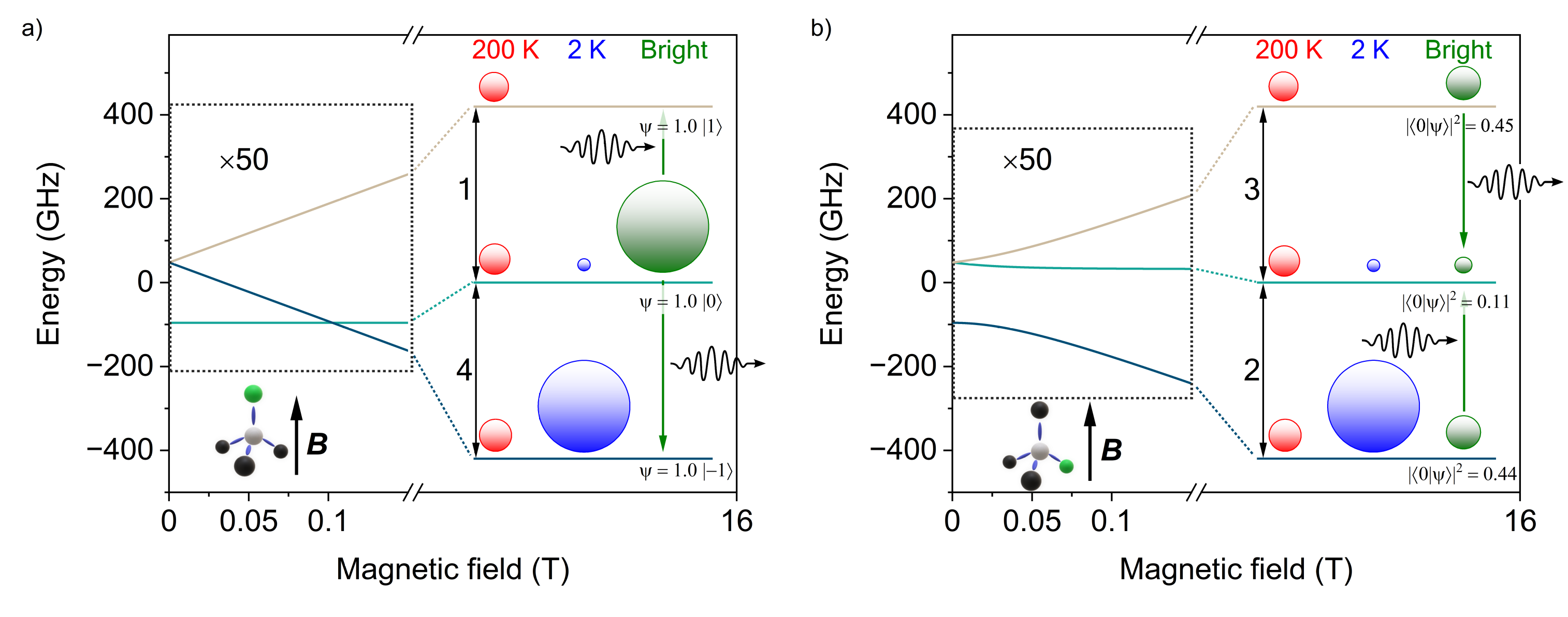} 
	\caption{\textbf{The magnetic field dependent energy level schemes when the magnetic field is parallel (a) or tetrahedral (b) to the NV direction}. The circle symbol sizes indicate the population of the respective levels for both dark (\textbf{{\color{red}200 K}} and \textbf{{\color{blue}2 K}}) and bright (\textbf{{\color{OliveGreen}Bright}}) conditions. The incident wavy arrow for the population under illumination indicates the net absorption of microwave photons and the outgoing indicates a net emission of such. The transition labeling is the same as used in Fig. 2. In the parallel case, the ZFS spin states are eigenstates. The full wave function of the tetrahedral case with complex weights is given in the SI in Eq. (2). As the population is set by the spin polarization mechanism preferring the $\ket{0}$ state, the square of the weight of $\ket{0}$ is given in (b).}
	\label{Fig3}
\end{figure*}

The LESR data unambiguously identify the light-sensitive transitions. Technical details are given in the Methods section and we only briefly discuss it here. The LESR method detects variations in the conventional ESR spectra under the action of an amplitude-modulated (or chopped) laser light using a double lock-in technique. The conventional ESR is detected with a magnetic field modulation technique using a lock-in amplifier at around $20$ kHz whereas the laser modulation frequency is three orders of magnitude lower (at around $20$ Hz). A second lock-in amplifier detects the variation in the conventional ESR in phase with the optical chopping. The output signal is thus proportional to the difference between the ESR spectra acquired with and without illumination. This is similar to the data shown in  Fig 1c. but the lock-in technique allows to measure much smaller changes in a phase-sensitive manner. While this method shows some resemblance to the more common optically detected magnetic resonance on NV centers \cite{Stepanov2015HFODMR}, no such experiments have been performed on this material to our knowledge, the reason probably being the lack of diamond samples with sufficiently large NV concentrations for this study.

Compared with Fig 1c., where the conventional ESR is shown with and without illumination, Fig 2b. shows the same effect more dramatically: (i) the increased intensity of the NV resonance at the lowest field value (Signal 1) appears as a (derivative) Lorentzian with a positive phase, (ii) whereas the decreased intensity of the highest NV resonance (Signal 4) translates as a line with negative phase in the LESR spectrum. These data reinforce the earlier observation of a light-induced emission for this signal. The P1 resonances of the neutral nitrogen atoms are missing from the LESR spectrum as they do not show optical activity. We performed similar studies in low-field ESR (at $\sim9.4$ GHz) and the key results are shown in the Supplementary Information.

However, somewhat surprisingly, we also observe that signals 1 and 2 as well as 3 and 4 behave similarly in a pairwise manner. We established above that signals 2 and 4 correspond to a lowest state to middle state transition, whereas 1 and 3 correspond to a middle state to the highest state transition. Clearly, the effect of population inversion and thus the emission is also reduced for signals 2 and 3 with respect to signals 1 and 4. 

These seemingly contradicting observations can be well explained by considering the actual level mixing for the different geometry as shown in Fig. 3. We show the energy level scheme for the two geometries, i.e., when the magnetic field is parallel to or at the tetrahedral angle to the NV-axis. For the earlier, the ZFS spin states remain eigenstates of the Zeeman Hamiltonian but a strong admixture is observed for the latter. The calculation shows that the transitions labeled as 2 and 4 correspond to the lowest to middle state transitions, which explains the temperature-dependent ESR data.

The magnitude of light-induced pumping can be obtained after calculating the eigenvectors in high magnetic field while using the original ZFS basis. When the magnetic field is parallel to the NV-axis, the middle state is populated as it has a purely $S_z=0$ character while the other two levels are the $S_z=\pm 1$ states. This explains why transition 4 shows emission, while transition 1 remains of the usual absorptive character. However, the effect is reversed for the other geometry: the lowest and highest states have in fact a \emph{stronger} $S_z=0$ character than that of the middle one as represented by the circle sizes in Fig. 3b. The three states are expressed in the ZFS basis in Eq. (2) in the SI and Fig. 3b shows the absolute square of the probability amplitude of the corresponding states being in the $\ket{0}$ state. The intensity of an ESR transition is proportional to the population difference between the final and initial states and the population is driven by the spin polarization preferring the $\ket{0}$ state. This means that upon light pumping, transition 3 has an emission character although weaker than that of transition 4.

A detailed calculation of the population of states at different temperatures and also under illumination is given in the Supplementary Information. The ESR intensity is proportional to the population difference of the corresponding states whereas the LESR signal is proportional to the change between the bright and dark population differences. The intensity ratios of LESR lines given by our model in $15~\text{T}$ external magnetic field read:
\begin{equation} \label{HFLESRintratios_main}
	I_1:I_2:I_3:I_4=0.94:0.27:-0.37:-1.00,
\end{equation}
where $I_2$ and $I_3$ refer to the intensity of one of the three possible tetrahedrally lying NV centers, i.e., in a perfect alignment the measured intensity should be three times this value due to the threefold degeneracy. However, in our case, there is a slight misalignment lifting the degeneracy, so the intensity values represented here refer to one ESR line (here the average of the three), not the sum of three. The phase of the observed LESR signal matches the predictions of the simplified model but the intensity ratio obtained from Fig. 2b. is:
\begin{equation} \label{HFLESRintratios_meas}
	I_1:I_2:I_3:I_4=0.58:0.09:-0.13:-1.00.
\end{equation}
It differs from the calculated values in Eq. \eqref{HFLESRintratios_main}, namely the inner lines are less intense than predicted by the wavefunctions. 

This is a manifestation of the polarization dependence of the electric dipole transition. The optical excitation of NV centers depends on the relative angle between the laser polarization and the NV center axis \cite{epstein2005anisotropic,alegre2007polarization}. In our experiment, the sample is placed with its $\langle 111\rangle$-axis parallel to the direction of propagation of the linearly polarized laser beam, leaving the other three NV directions at a tetrahedral angle with respect to it. As shown in Ref. \cite{alegre2007polarization} in a similar geometry, the pumping rate ($P$) of the NV centers lying at a tetrahedral angle strongly depends on the angle ($\varphi$) between the laser polarization and the projection of the NV center on the surface. The effective pumping rate is:
\begin{equation}\label{eq:polarization}
        P=P_0  \left(\text{sin}^2\varphi+\frac{1}{9}\text{cos}^2\varphi\right),
\end{equation}
where $P_0$ is the pumping rate of the NV centers in the $\langle 111\rangle$ orientation. Depending on the rotation angle around the $\langle 111\rangle$-axis, this factor varies between $1$ and $1/9$ continuously. However, for the three possible orientations denoted by angles $\varphi$, $\varphi-120^\circ$ and $\varphi+120^\circ$ the sum ($P_{\varphi}+P_{\varphi-120^\circ}+P_{\varphi+120^\circ}$) is always $\frac{5}{3}P_0$. With this correction, the theoretical values of $I_2$ and $I_3$ are $0.15$ and $-0.2$, respectively. To explain the remaining difference and the experimentally lower value of $I_1$, the insufficient optical power (due to the sample optical density mentioned earlier) should be included in a more sophisticated model.

Aside from the better sensitivity to detect the optically induced changes in ESR, the LESR technique also enables studying spin dynamics. Knowledge of the relevant ESR relaxation times ($T_1$ and $T_2$) in the NV centers plays a key role in applications, e.g., quantum technology, metrology, sensing, and the maser applications \cite{Rondin2014magnetoReview, Ladd2010quantumcomputers, Breeze2018}. These relaxation times are usually measured using pulsed ESR techniques. However, such instruments are not yet available at high frequencies. It would also be possible to measure the relatively long spin-lattice relaxation time, $T_1$, in NV centers using saturation experiments \cite{SlichterBook,Markus2023NatCommun} but the near-THz ESR instruments yet lack the required absolute power accuracy.

In the LESR technique, $T_1$ is measurable by staying on a given resonance line (we studied Signals 1 and 4, i.e., when the magnetic field is parallel to the NV-axis) and varying the frequency of the optical chopping in the $1-400~\text{Hz}$ range. During irradiation, the level populations start deviating from their equilibrium values; when the light is switched off, the system recovers to equilibrium with a time constant of $T_1$.

\begin{figure}[!h]
    \centering
	\includegraphics*[width=1\linewidth]{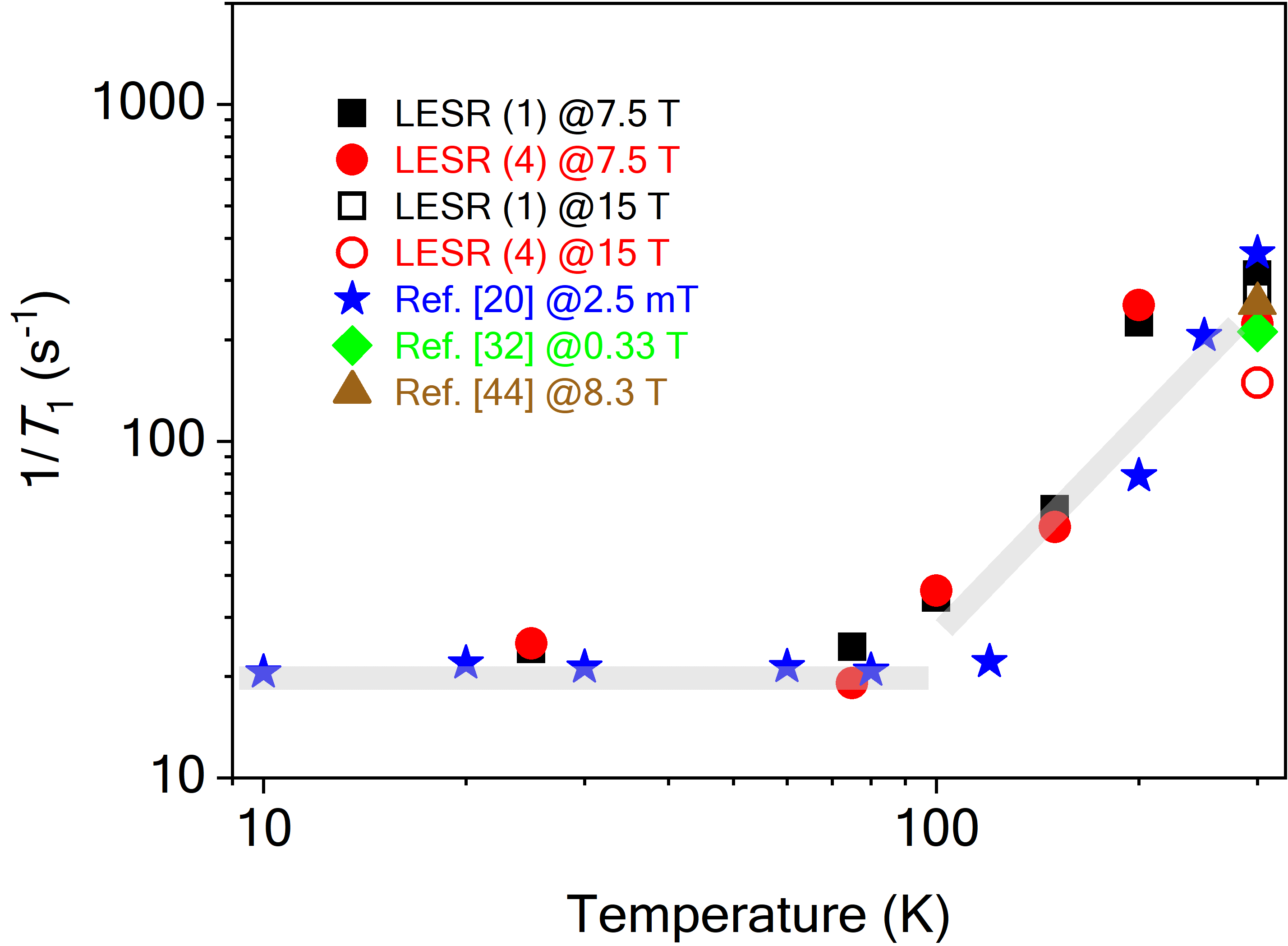} 
	\caption{{\color{black}\textbf{Temperature dependence of spin-lattice relaxation time of NV centers measured by LESR technique.}} $1/\text{$T_1$}$ as a function of temperature measured on the lowermost (\textbf{LESR (1)}) and uppermost lying (\textbf{{\color{red}LESR (4)}}) NV center LESR lines at 7.5 T (filled symbols) and also at room temperature for 15 T (open symbols). LESR (1) and (4) correspond to Signal 1 and 4 of Fig. 2, respectively. For comparison, we show previously reported data on diamond samples with similar NV center density: a full temperature dependence at \textbf{{\color{blue}2.5 mT}} \cite{Jarmola2012} and two room temperature results at \textbf{{\color{green}0.33 T}} \cite{Kollarics2022} and at \textbf{{\color{darkgoldenrod}8.3 T}} \cite{fortman2021electron}. Gray lines are guide to the eye indicating a residual relaxation and a phonon-dominated region.}
	\label{Fig4}
\end{figure}

In Fig 4., $1/T_1$ values obtained with the double modulation technique are presented and further technical details are given in the SI. Temperature-dependent data is given at $7.5$ T and data at room temperature is shown for $15$ T ({\color{black} $0.42\ \text{THz}$}). The lower overall sensitivity of the HFESR spectrometer at {\color{black} $0.42\ \text{THz}$} (it is optimized for {\color{black} $0.21\ \text{THz}$} \cite{Nafradi2008a,Nafradi2008b}) combined with the extensive measurement time and magnetic field instability limited these measurements further. We include in the figure previously reported experimental data measured on samples that are similar to ours, i.e., HPHT grown, exposed to high-dose electron beam irradiation, resulting in a concentration above $10~\text{ppm}$ of NV centers after high-temperature annealing \cite{Jarmola2012}. 

\section*{Discussion}

Our temperature-dependent $T_1$ results agree remarkably well with the results in Ref. \cite{Jarmola2012}, although the latter experiment was performed at $B=2.5~\text{mT}$, i.e., in a $6,000$ times smaller magnetic field. Also, pulsed X-band (at $0.33$ T) ESR measurements \cite{Kollarics2022} show a $T_1$ relaxation time of about $4.73~\text{ms}$ at room temperature while high-field optically detected magnetic resonance measurements yield $3.9~\text{ms}$ \cite{fortman2021electron}. These results are in good agreement with our high field LESR results at $7.5~\text{T}$ and at $15~\text{T}$. This indicates that the spin-lattice relaxation is unaffected by the high magnetic fields, which is important for the future use of the NV center in the aforementioned applications. 

{\color{black}Our measurement does not directly determine the $T_2$ (spin decoherence time) and $T_2^*$ (the spin dephasing time) albeit both are relevant for spintronics applications, especially to those where the interplay between coupled spin-polarized carriers and photons is exploited \cite{coherentspinprecession,spinlasers}. We nevertheless do not observe a significant line broadening although the applied magnetic field is 50 times larger than the customary $0.3\ \text{T}$ in X-band. This is expected for $T_2$ as it is caused by spin-spin interactions (both like and unlike spin-spin interactions), described by the van Vleck formula \cite{AbragamBook}. The $T_2^*$ could in principle be caused by magnetic field inhomogeneity but for solid-state systems, the dephasing time is dominated by impurities or by unresolved hyperfine interactions \cite{Portis}, both of which are independent of the magnetic field. We therefore expect that both $T_2$ and $T_2^*$ can be well estimated by pulsed ESR experiments in low fields, where such measurements are possible \cite{Kollarics2022}. This is very promising for the expected TASER applications as the cooperativity factor would be limited by a short $T_2^*$.}

A conveniently long $T_1$ is required to preserve a high level of population inversion as otherwise this process would deplete the inverted population. This potentially makes the diamond NV center in high magnetic field a possible candidate for a coherent THz source (a TASER) or THz amplifier with ample technical possibility for magnetic field tuning and modulation, which may be eventually advantageus for conventional and quantum communication and various condensed matter applications. {\color{black}We outline and critically assess the possible construction of a diamond NV based TASER in the SI.} However, exploiting the true capabilities of this system requires substantial additional work to properly describe the TASER relevant cooperativity factor \cite{Breeze2018} in high magnetic field.

In summary, we showed that the diamond NV centers can operate as sources of coherent THz emission when subjected to a high magnetic field. This effect is achieved by pumping the sample with visible light and achieving a population inversion. The dependence of the pumping efficiency on the respective orientation of the NV-axis and the magnetic field was observed using a sensitive method, light-induced ESR. The observation was explained by considering the spin Hamiltonian of the NV center and also by the variation of the light polarization with respect to the NV-axis. The LESR technique was also used to determine the spin-lattice relaxation time, $T_1$ at $7.5~\text{T}$ and at $15~\text{T}$, and it showed that the dominant relaxation mechanisms are unaffected by the magnetic field. 

\section*{Materials and Methods}

All experiments presented in this paper were carried out on a single crystal diamond plate (Type 1b produced by high-pressure/high-temperature method, supplied by Element Six Ltd.) of a hexagonally-cut shape (diameter about $5~\text{mm}$) with NV center concentration of about {\color{black}$12~\text{ppm}$} which was produced by electron beam irradiation and subsequent thermal annealing. More details on sample preparation and quantitative analysis are given in Ref. \cite{Kollarics2022} {\color{black}and in the SI}. The block diagram of the high-frequency light-induced electron spin resonance spectrometer is shown in Fig. 5. It is based on a high-frequency electron spin resonance (HFESR) spectrometer operating in the {\color{black}$0.052-0.420~\text{THz}$} frequency range is described in Ref. \cite{Nafradi2008a}. This instrument works with phase-locked loop stabilized, frequency-multiplied microwave (of near-THz) sources at a number of discrete frequencies. Detection is based on a liquid helium cooled InSb hot-electron bolometer used in a homodyne single-ended mixer configuration. The detector is part of a millimeter-wave, quasi-optical bridge, which allows for the phase-sensitive detection of reflected radiation from the sample. 

\begin{figure}[!ht]
	\centering
	\includegraphics*[width=1\linewidth]{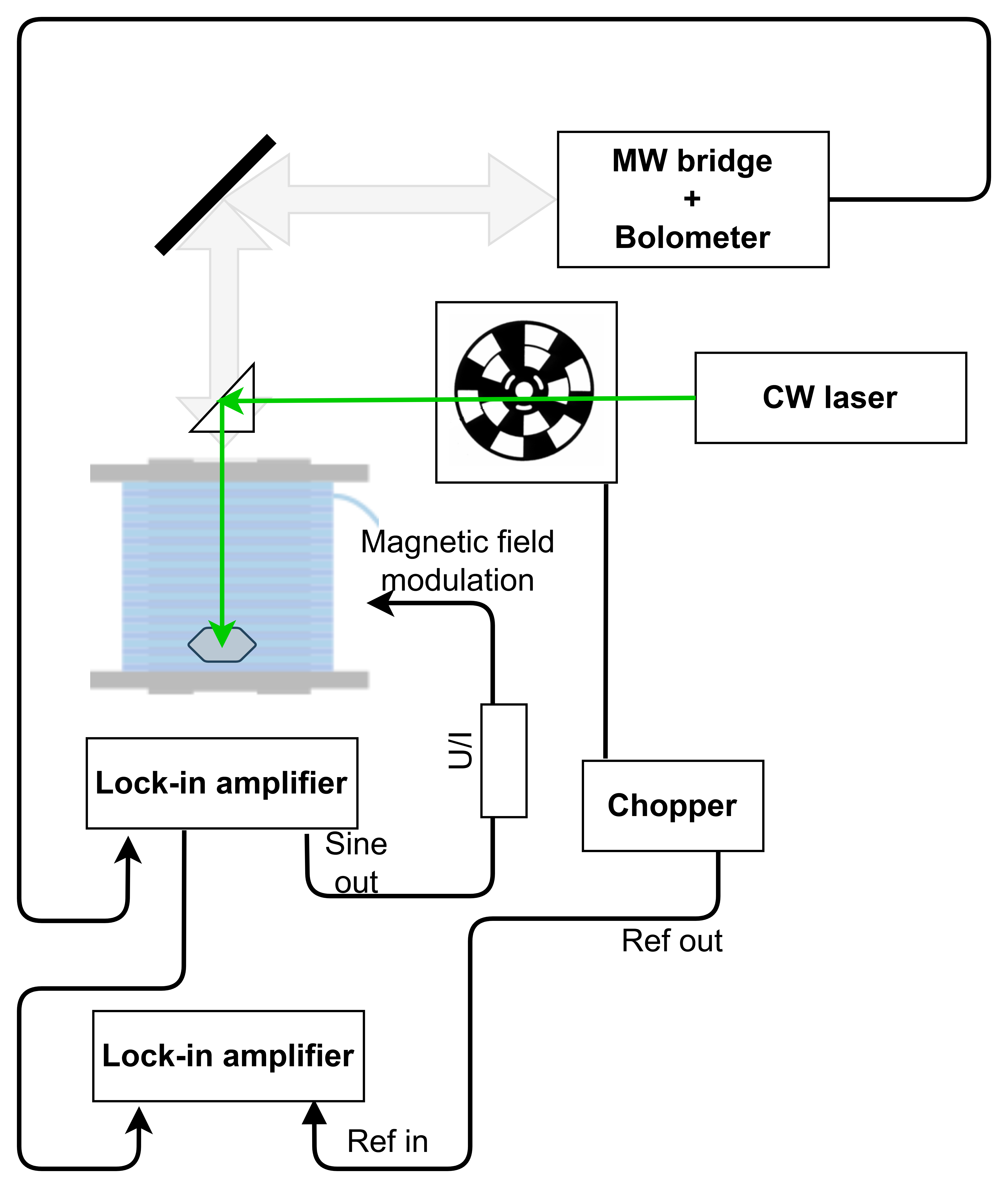}
	\caption{\textbf{High-field/high-frequency ESR setup with optical excitation}. Microwaves are directed through a corrugated waveguide and the reflected waves are detected using a bolometer. The signal from the bolometer is fed into a lock-in amplifier which modulates the external magnetic field. Its down-mixed output is fed into a second lock-in amplifier, controlling the visible laser chopping, eventually producing the light-induced ESR signal.}
	\label{setup}
\end{figure}

The sample is placed on a piezo-controlled single-axis goniometer (attocube systems AG) and is inside a liquid helium VTI inside a superconducting solenoid with fields up to $16~\text{T}$ (Oxford Instruments). The sample chamber is isolated from the VTI and is embedded in helium exchange gas. Microwaves propagate back and forth to the sample from the top in a corrugated waveguide which is covered by a $1"$ clear diameter polyethylene window. Optical access to the sample was also obtained from the top. The original PE window was first replaced by a $5~\text{mm}$ thick BK7 glass but it caused more than $50\%$ decrease in microwave power reaching the sample. We found that a thin ($\approx 300~\mu\text{m}$) sheet of Mylar (biaxially-oriented polyethylene terephthalate) is a good compromise, excellent for vacuum isolation and it causes less than $10\%$ microwave loss for our frequency range, while it is transparent in the visible-light range resulting in only $5-10\%$ of optical power lost.

{\color{black}The sample is at the bottom of a 2-meter-long corrugated waveguide that efficiently directs the millimeter waves toward the sample. The waveguide ends in a hollow brass tube with an approximate diameter of $5\ \text{mm}$. The system is equipped with a frequency-doubled Nd:YAG laser working at $\lambda = 532\ \text{nm}$ with a maximum output power of $150\ \text{mW}$. Guiding the light was achieved by constructing a precise optics breadboard with four mirrors and pinholes, which were fixed to the probehead. The light passes through a Mylar window that introduces a $5-10\%$ loss in the visible spectrum. The main loss in optical power reaching the sample is caused by the beam divergence and the resulting reflections from the walls of the waveguide. After careful alignment of the mirrors, the maximum power passing through the brass tube was $20\ \text{mW}$. This incident power equals to an intensity of $1\ \text{W/}\text{cm}^2$.}

A magnetic field, parallel to the DC magnetic field, is modulated with an amplitude of $0.05~\text{mT}$ at $20~\text{kHz}$, which is customary in ESR spectroscopy. The bolometer output signal is detected phase-sensitively with a lock-in amplifier (Stanford Research Systems, SR830). This output provides the conventional HFESR signal. To obtain the light-induced ESR signal, the laser light is modulated with an optical chopper (Stanford Research Systems, SR540). Then, the demodulated output signal from the first lock-in amplifier is fed into a second lock-in amplifier (SR830) which controls the chopper. The second demodulated output produces the light-induced ESR signal as demonstrated in Refs. \cite{Lane1992,ShinarPRL2005}.



\cite{Thiering2018,goree1985doubleLI,subedi,savvin,wee2007cross-section,chapman,Man_2016,Rose_2018,Reiserer_2016,Thiering_2017,Tetienne_2012,Goldman_2015,Goldman_2015_PRB,microcavity,rogalin2018optical}



\section*{Acknowledgments}
\textbf{Acknowledgments:} R.~Ga\'al is acknowledged for his technical expertise and help, A. J\'anossy and A. Sienkiewicz are acknowledged for enlightening discussions and for kind assistance with the experimental setup. \textbf{Funding:} This work was supported by the Hungarian National Research, Development and Innovation Office (NKFIH) Grants 2022-2.1.1-NL-2022-00004, K137852, TKP 2021-NVA-04, and the V4-Japan Joint Research Program (BGapEng, 2019-2.1.7-ERA-NET-2021-00028). A.G. acknowledges the NKFIH grant no. KKP129866 of the National Excellence Program of the Quantum-Coherent Materials Project, the EU QuantERA II MAESTRO project, and the Horizon Europe EIC Pathfinder QuMicro project (grant no. 101046911).
\textbf{Author Contributions}: All authors discussed the results and approved the manuscript. S. K. and S. F. conceived the project. S. K., B. G. M., R. K., G. N., K. K., and F. S. carried out the experimental work. S. K., G. T., \'{A}. G., L. F., and F. S. analyzed the data and supported the results with theoretical work. L. F. and F. S. acquired the funding and supervised the research. \textbf{Competing interests:} All authors declare that they have no competing interests.
\textbf{Data Availability:} All data needed to evaluate the conclusions in the paper are present in the paper and/or the Supplementary Materials.


\clearpage
\begin{center}
	\Large{Supplementary material to: Terahertz Emission from Diamond Nitrogen-Vacancy Centers}
\end{center}
\normalsize


{\color{black}\section{Sample preparation}
	Type 1b single crystal diamond produced by the high pressure high temperature method was purchased from Element Six Ltd (Didcot, UK). The crystal was exposed to electron irradiation in a radio frequency linear accelerator with variable energy between $1\ \text{MeV}$ and $4\ \text{MeV}$ built by RadiaBeam Technologies LLC (Santa Monica, CA, USA). The sample received a total electron fluence of $2.18\cdot10^{18}\ 1/\text{cm}^2$ determined from the resulting current of the accelerator. Subsequently, the sample was annealed at $800\ ^\circ\text{C}$ for two hours and then at $1000\ ^\circ\text{C}$ for two hours again. NV center concentration was determined first relative to nitrogen concentration and then calibrated against copper(II) sulfate pentahydrate, $\ce{CuSO4}\cdot 5\ce{H2O}$ standard in a conventional X-band continuous wave electron spin resonance (CW ESR) spectrometer (Bruker Elexsys E500). The ratio of NV centers compared to neutral substitutional nitrogen atoms in the diamond lattice was found to be $13.2\%$ yielding $12\ \text{ppm}$ NV concentration considering $90\ \text{ppm}$ nitrogen concentration based on the CW ESR spectra. 
}
\section{Population of states in high and low magnetic field}

In this section, we show the calculation behind the statements in the main text related to the population of the eigenstates in a finite magnetic field. We diagonalize the spin Hamiltonian in two cases: (i) when the external magnetic field $\boldsymbol{B}$ is parallel with the NV-axis and (ii) when $\boldsymbol{B}$ forms a tetrahedral angle with the NV-axis, i.e., the orientation of the other three NV-s in case (i). The resulting energies are expressed in temperature ($\text{K}$) for convenience. Without illumination, the population is described by the Boltzmann distribution, where the probability of a particle occupying state \emph{i} is:
\begin{equation}\label{probabilities}
	p_i=\frac{N_i}{N}=\frac{\exp(-\frac{\varepsilon_i}{\kB T})}{\sum_{i}^{} \exp(-\frac{\varepsilon_i}{\kB T})}.
\end{equation}

The LESR experiments presented in Fig. 2 in the main text were carried out at $200~\text{K}$ and in $15~\text{T}$ so we use these values in our present calculation. During illumination, the spin polarization effect competes with the Boltzmann distribution. The spin polarization is described with a set of differential equations (rate equations) introducing several transition probabilities \cite{Thiering2018}. In high magnetic field and at a finite temperature, the exact populations could only be given with a system of differential equations using additional input parameters such as absorbed optical power and spin-lattice relaxation time aside from the already complicated spin polarization dynamics. Our aim is to understand the light-induced changes to the ESR signal in high magnetic fields and to qualitatively explain the LESR signal presented in Fig. 2 in the main text. For that, we investigate the high optical intensity regime where the effect of finite temperature can be neglected. So in our simplified model, the population of a given state is proportional to the $S_z=0$ coefficient. In the parallel case, the finite magnetic field eigenstates are pure $S_z$ states therefore when illuminated the middle state is fully populated whereas in a tetrahedral case, the population depends on the strength of the magnetic field. Whereas in the tetrahedral case, the eigenstates are linear combinations of the pure ZFS states such as shown here in Eq. \eqref{eq:brightwavefunctions} in an external magnetic field of $15~\text{T}$:
\begin{equation}\label{eq:brightwavefunctions}
	\begin{split}
		\Psi_{\mathrm{U}}=(-0.5761-0.3326\,\cplxi)\ket{-1}+(0.5786+0.3340\,\cplxi))\ket{0}\\+(0.2887+0.1667\,\cplxi)\ket{1},\\
		\Psi_{\mathrm{M}}=(0.1729-0.6453\,\cplxi)\ket{-1}+(0.0863-0.3220\,\cplxi)\ket{0}\\+(0.1722-0.6426\,\cplxi)\ket{1},\\
		\Psi_{\mathrm{L}}=0.3333\ket{-1}+0.6652\ket{0}-0.6681\ket{1}.
	\end{split} 
\end{equation}

Diagonalizing the spin Hamiltonian in $15~\text{T}$ with ZFS of $0.1~\text{T}$ we acquire eigenvalues of the lowest, middle and upper states $-20.14~\text{K,}~-0.09~\text{K}~\text{and } 20.23~\text{K}$ in the parallel and $-20.20~\text{K,}~0.03~\text{K}~\text{and } 20.17~\text{K.}$ in the tetrahedral orientation. Substituting this into equation (\ref{probabilities}), we can calculate the population of the given states.

\begin{table}[!ht]
	\centering
	\resizebox{\columnwidth}{!}{
		\begin{tabular}{||c | c | c | c | c ||} 
			\hline
			\thead{$B=15~\text{T}$ \\ parallel case} & \multicolumn{3}{c|}{Temperature} &\\
			\hline
			State & $2~\text{K}$ & $20~\text{K}$ & $200~\text{K}$ & Bright \\
			\hline
			U & $0$ & $0.08876$ & $0.30024$ & $0$ \\
			\hline
			M & $0.00004$ & $0.24299$ & $0.33033$ & $1$\\
			\hline
			L & $0.99996$ & $0.66825$ & $0.36740$ & $0$\\
			\hline
			\hline
			\thead{$B=15~\text{T}$ \\ tetrahedral case} & \multicolumn{3}{c|}{Temperature} &\\
			\hline
			State & $2~\text{K}$ & $20~\text{K}$ & $200~\text{K}$ & Bright \\
			\hline
			U & $0$ & $0.08876$ & $0.30033$ & $0.46$ \\
			\hline
			M & $0.00004$ & $0.24299$ & $0.33215$ & $0.11$\\
			\hline
			L & $0.99996$ & $0.66825$ & $0.36751$ & $0.43$\\
			\hline
	\end{tabular}}
	\caption{\textbf{Population of states at given temperatures and under illumination in $\boldsymbol{15}~\text{T}$.}}
	\label{table:15Tpop}
\end{table}

\begin{table}[!ht]
	\centering
	\resizebox{\columnwidth}{!}{
		\begin{tabular}{||c | c | c | c | c | c ||} 
			\hline
			\thead{$B=15~\text{T}$ \\ parallel case} & \multicolumn{3}{c|}{Temperature} & & \\
			\hline
			State & $2~\text{K}$ & $20~\text{K}$ & $200~\text{K}$ & Bright & Bright-Dark\\
			\hline
			M-U & $0$ & $0.1561$ & $0.0321$ & $1$ & $0.9679$\\
			\hline
			L-M & $0.9999$ & $0.4221$ & $0.0350$ & $-1$ & $-1.035$\\
			\hline
			\hline
			\thead{$B=15~\text{T}$ \\ tetrahedral case} & \multicolumn{3}{c|}{Temperature} & & \\
			\hline
			State & $2~\text{K}$ & $20~\text{K}$ & $200~\text{K}$ & Bright & Bright-Dark \\
			\hline
			M-U & $0$ & $0.1542$ & $0.0318$ & $-0.35$ & $-0.3818$ \\
			\hline
			L-M & $0.9999$ & $0.4253$ & $0.0354$ & $0.32$ & $0.2846$ \\
			\hline
	\end{tabular}}
	\caption{\textbf{Population difference between states at given temperatures and under illumination in $\boldsymbol{15}~\text{T}$.}}
	\label{table:15Tdiff}
\end{table}

The ESR intensity is proportional to the population difference. Therefore the CW ESR intensity ratios of NV centers in $15~\text{T}$ in the strong illumination limit would yield:
\begin{equation} \label{HFESRintratios}
	I_1:I_2:I_3:I_4=1:0.32:-0.35:-1
\end{equation}

The LESR measures the change induced by the light, therefore the intensity is proportional to the difference in the population difference in the bright and dark measurements:
\begin{equation} \label{HFLESRintratios}
	I_1:I_2:I_3:I_4=0.9679:0.2846:-0.3818:-1.035
\end{equation}

Diagonalizing the spin Hamiltonian in $0.35~\text{T}$ with ZFS of $0.1~\text{T}$ we acquire eigenvalues of the lowest, middle and upper states $-0.48~\text{K,}~0.001~\text{K}~\text{and } 0.47~\text{K}$ in the parallel and $-0.49~\text{K,}~0.03~\text{K}~\text{and } 0.46~\text{K}$ in the tetrahedral orientation. Substituting this into equation (\ref{probabilities}), we can calculate the population of the given states.

\begin{table}[!ht]
	\centering
	\resizebox{\columnwidth}{!}{
		\begin{tabular}{||c | c | c ||} 
			\hline
			\thead{$B=0.35~\text{T}$ \\ parallel case} & \thead{Temperature \\ $300\ \text{K}$} & \\
			\hline
			State & Dark & Bright \\
			\hline
			U & $0.3328$ & $0$ \\
			\hline
			M & $0.3333$ & $1$\\
			\hline
			L & $0.3339$ & $0$\\
			\hline
			\hline
			\thead{$B=0.35~\text{T}$ \\ tetrahedral case} & \thead{Temperature \\ $300~\text{K}$} & \\
			\hline
			State & Dark & Bright \\
			\hline
			U & $0.3328$ & $0.5314$ \\
			\hline
			M & $0.3333$ & $0.1084$\\
			\hline
			L & $0.3339$ & $0.3602$\\
			\hline
	\end{tabular}}
	\caption{\textbf{Population of states at room temperature in $\boldsymbol{0.35}~\text{T}$.}}
	\label{table:Xbandpop}
\end{table}

\begin{table}[!ht]
	\centering
	\resizebox{\columnwidth}{!}{
		\begin{tabular}{||c | c | c | c ||} 
			\hline
			\thead{$B=0.35~\text{T}$ \\ parallel case} & \thead{Temperature \\ $300~\text{K}$} & & \\
			\hline
			State & Dark & Bright & Bright-Dark \\
			\hline
			M-U & $0.0005$ & $1$ & $0.9995$ \\
			\hline
			L-M & $0.0006$ & $-1$ & $-1.0006$ \\
			\hline
			\hline
			\thead{$B=0.35~\text{T}$ \\ tetrahedral case} & \thead{Temperature \\ $300~\text{K}$} & &  \\
			\hline
			State & Dark & Bright & Bright-Dark \\
			\hline
			M-U & $0.0005$ & $-0.4235$ & $-0.4235$ \\
			\hline
			L-M & $0.0006$ & $0.2513$ & $0.2507$ \\
			\hline
	\end{tabular}}
	\caption{\textbf{Population difference at room temperature in $\boldsymbol{0.35}~\text{T}$.}}
	\label{table:Xbanddiff}
\end{table}

The CW ESR intensity ratios of NV centers in $0.355~\text{T}$ in the strong illumination limit:
\begin{equation} \label{ESRintratios}
	I_1:I_2:I_3:I_4=1:0.25:-0.42:-1
\end{equation}

The LESR intensity ratios in $0.355~\text{T}$ in the strong illumination limit:
\begin{equation} \label{LESRintratios}
	I_1:I_2:I_3:I_4=0.9995:0.2494:-0.4205:-1.0006
\end{equation}

\section{Dark and bright room temperature CW ESR spectra}

\begin{figure}[!ht]
	\centering
	\includegraphics*[width=1\linewidth]{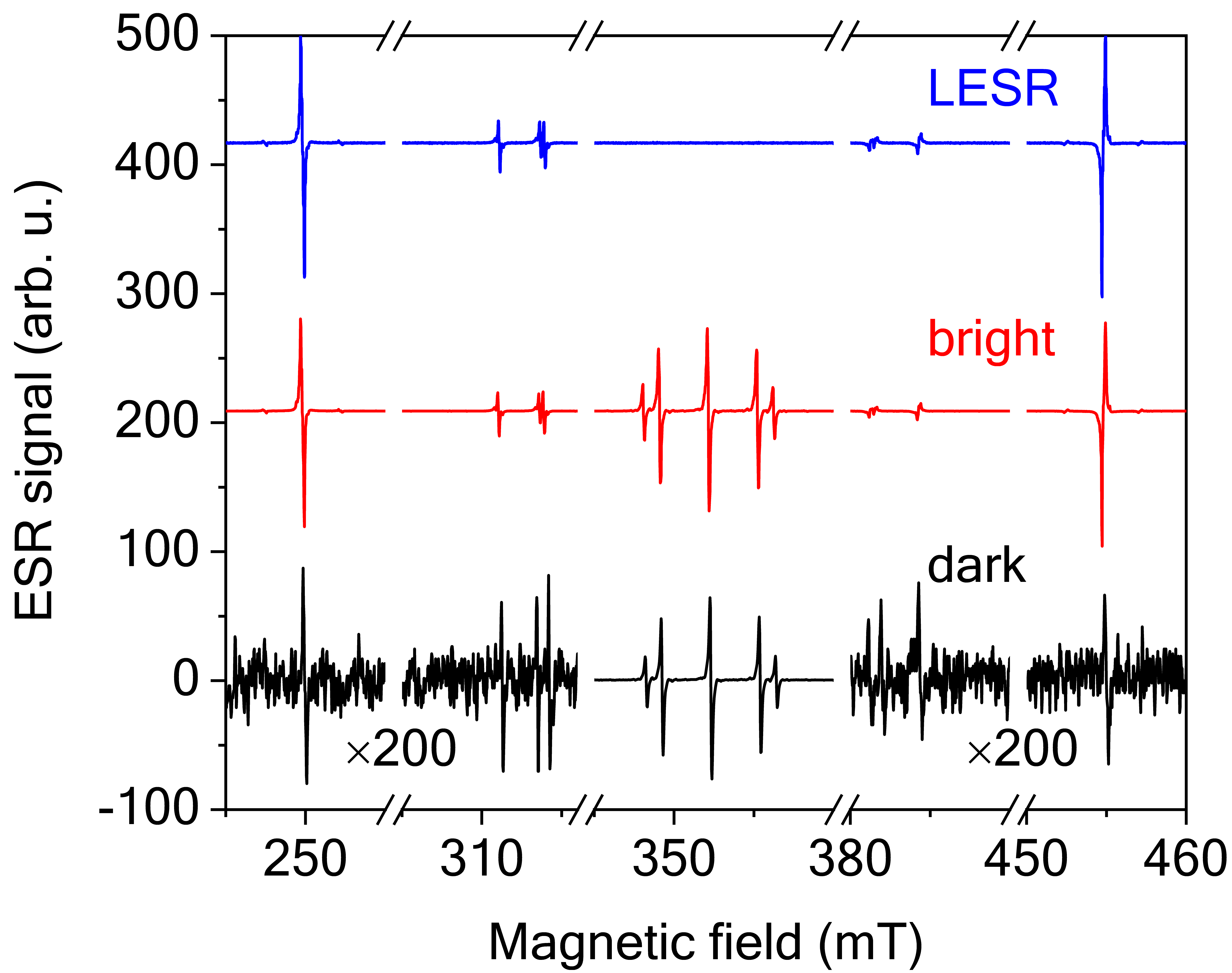}
	\caption{\textbf{X-band ($\boldsymbol{f\approx} \boldsymbol{9.8}~\text{GHz}$) light-induced ESR}. Resonances of NV centers are scaled up with a factor of 200 for clarity in the spectrum acquired without illumination (black line). Under continuous illumination, the lower four resonances grow in intensity and the upper four resonances reverse sign as a clear indication of population inversion.}
	\label{SM1}
\end{figure}

In Fig. S1 room temperature X-band ($B_0 \approx 0.35~\text{T}$) ESR spectra are shown measured with and without illuminating the diamond sample. In the dark spectrum, the part with resonant lines of NV centers is scaled up with a factor of $200$, whereas the bright spectrum is the raw data. The data presented in Fig. S1 is in qualitative accordance with the intensity ratios calculated in Eq. \eqref{ESRintratios}, namely the sign of the intensities resembles the sign of the resonances. The model predicts less intense inner lines but the measurement shows even smaller results. 

\section{Determining $T_1$ from the double-modulated LESR signal}

Without illuminating the sample, the spin population of the system is given by the Boltzmann equilibrium, while illumination leads to the spin-polarized steady-state. Switching the illumination on and off, the system starts to relax to its proper steady-state enabled by the spin-lattice relaxation. Therefore, the input voltage of the second lock-in amplifier follows a single exponential decay with the spin-lattice relaxation time as time constant. The phase-sensitive detection of the second lock-in can be mathematically described as multiplying the input voltage with the reference signal (with local oscillator frequency of $f=2\pi\omega$) and integrating it over a full period:
\begin{equation} \label{eq:lockindecay_cos}
	\begin{split}
		X= \int_{0}^{\frac{2\pi}{\omega}} \textnormal{cos}(\omega t) \expe^{-t/T_1} \,\dd t = T_1 \frac{1}{1+\omega^2 T_1^2} \left(1-\expe^{\frac{2\pi}{\omega T_1}}\right)
	\end{split}
\end{equation}
\begin{equation} \label{eq:lockindecay_sin}
	\begin{split}
		Y= \int_{0}^{\frac{2\pi}{\omega}} \textnormal{sin}(\omega t) \expe^{-t/T_1} \,\dd t = T_1 \frac{\omega T_1}{1+\omega^2 T_1^2} \left(1-\expe^{\frac{2\pi}{\omega T_1}}\right).
	\end{split}
\end{equation}
Consequently, the magnitude of the LESR signal:
\begin{equation} \label{eq:magnitude}
	\begin{split}
		R & =\sqrt{X^2+Y^2} \\ & =T_1 \left(1-\expe^{\frac{2\pi}{\omega T_1}}\right) \sqrt{\left(\frac{1}{1+\omega^2 T_1^2}\right)^2+\left(\frac{\omega T_1}{1+\omega^2 T_1^2}\right)^2} \\
		& =T_1 \left(1-\expe^{\frac{2\pi}{\omega T_1}}\right) \frac{1}{\sqrt{1+\omega^2 T_1^2}}.
	\end{split}
\end{equation}
Therefore, by sweeping the modulation frequency and fitting the acquired LESR magnitude with
\begin{equation} \label{eq:lowpass}
	S(f)= \alpha \left(1-\expe^{\frac{1}{f \beta}}\right) \frac{1}{\sqrt{1+\left(2\pi f\right)^2 \beta^2}}
\end{equation}
one yields the spin-lattice relaxation time as $\beta$ as shown in Fig. S2. The parameter $\alpha$ is proportional to $T_1$ but also to the steady-state magnetization. Note that in the above derivation, $\alpha=T_1$ holds only because the amplitude of the input signal in Eq. \eqref{eq:lockindecay_cos} and in Eq. \eqref{eq:lockindecay_sin} is set to one.

\begin{figure}[!ht]
	\centering
	\includegraphics*[width=1\linewidth]{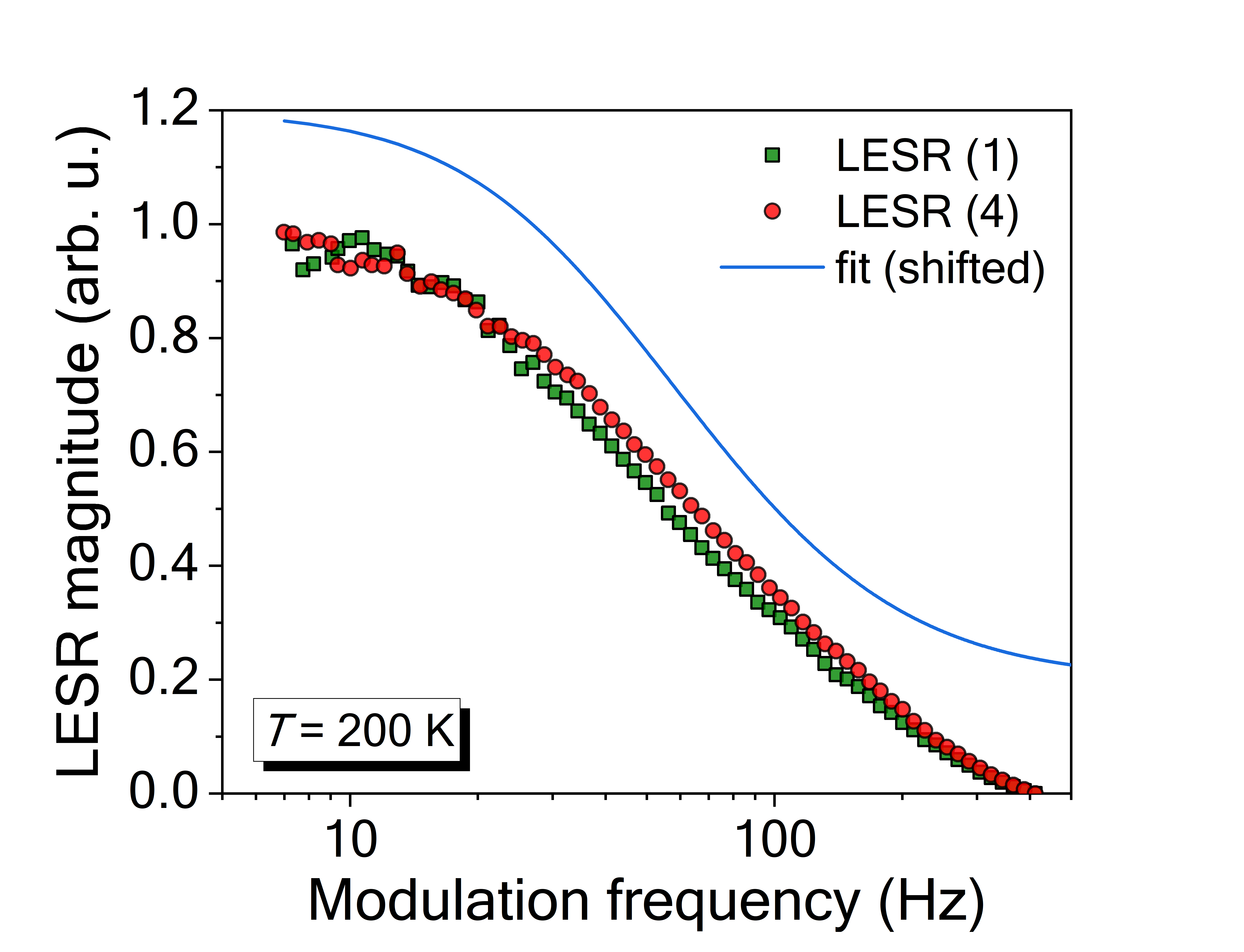}
	\caption{\textbf{LESR magnitude of Signal 1 ({\color{OliveGreen}green squares}) and 4 ({\color{red}red circles}) as a function of the chopping modulation frequency that was swept in the $\boldsymbol{1-400}~\text{Hz}$ range at $\boldsymbol{200}~\text{K}$}. The fitted low-pass characteristic, described by Eq. \eqref{eq:lowpass}, is shown with a straight blue line, shifted for better visibility.}
	\label{fig:lowpass}
\end{figure}

\section{Additional details of the LESR technique}

\begin{figure}[!htb]
	\centering
	\includegraphics*[width=1\linewidth]{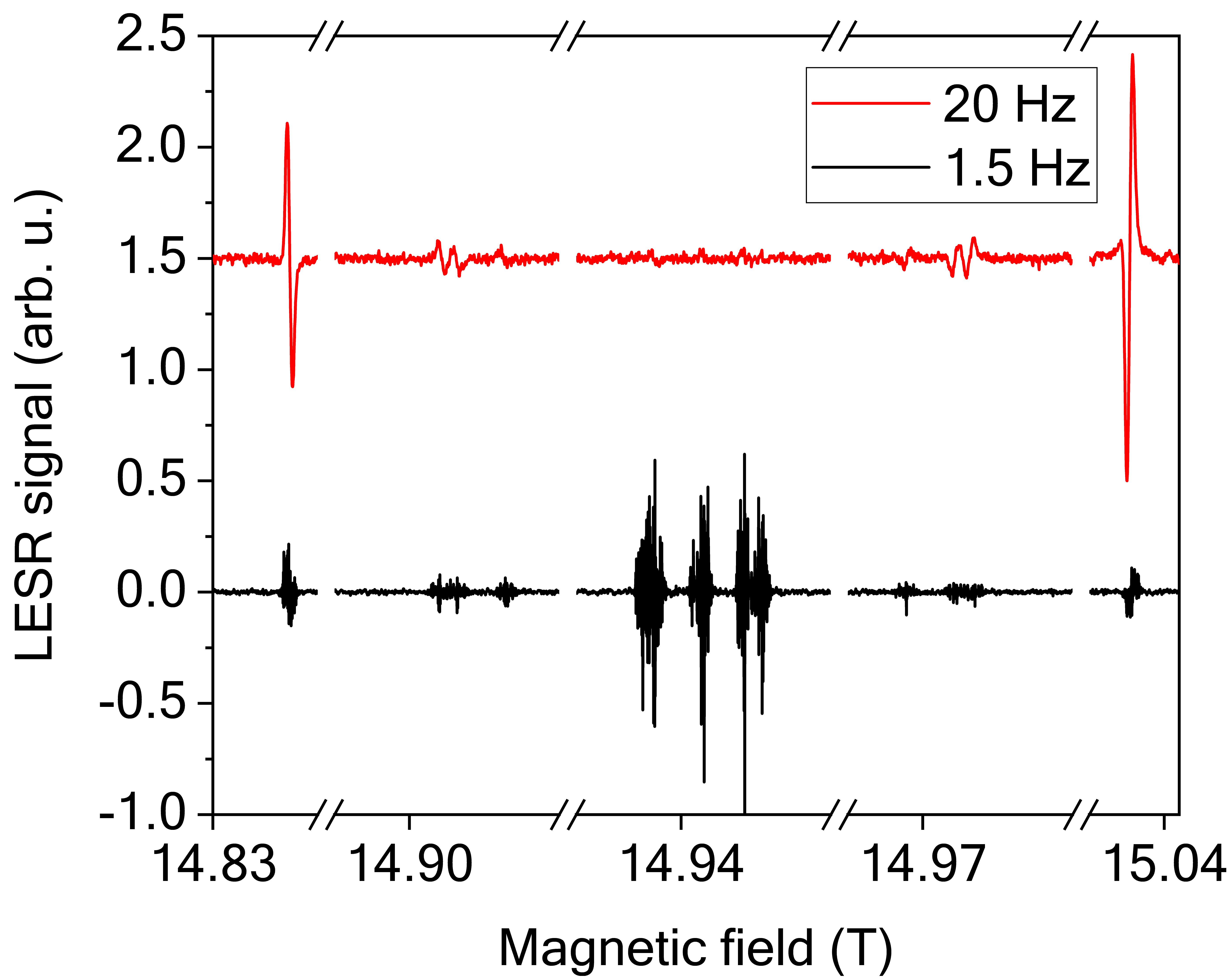}
	\caption{\textbf{Light-induced ESR signal of NV centers}. Both spectra were acquired with the same magnetic field sweep velocity of $0.004~\text{T}/\text{min}$ but with different optical modulation frequencies.}
	\label{varying_chop}
\end{figure}

{\color{black}Figure} S3 shows LESR spectra acquired with the same magnetic field sweep velocity but with two different optical modulation frequencies. In both cases, the optical modulation is far from the $20~\text{kHz}$ magnetic field modulation as the principles of double modulation technique require it \cite{goree1985doubleLI}, but the second modulation frequency (the optical chopping) interferes with the sweeping resulting in spectral leakage.

\begin{figure}[!htb]
	\centering
	\includegraphics*[width=1\linewidth]{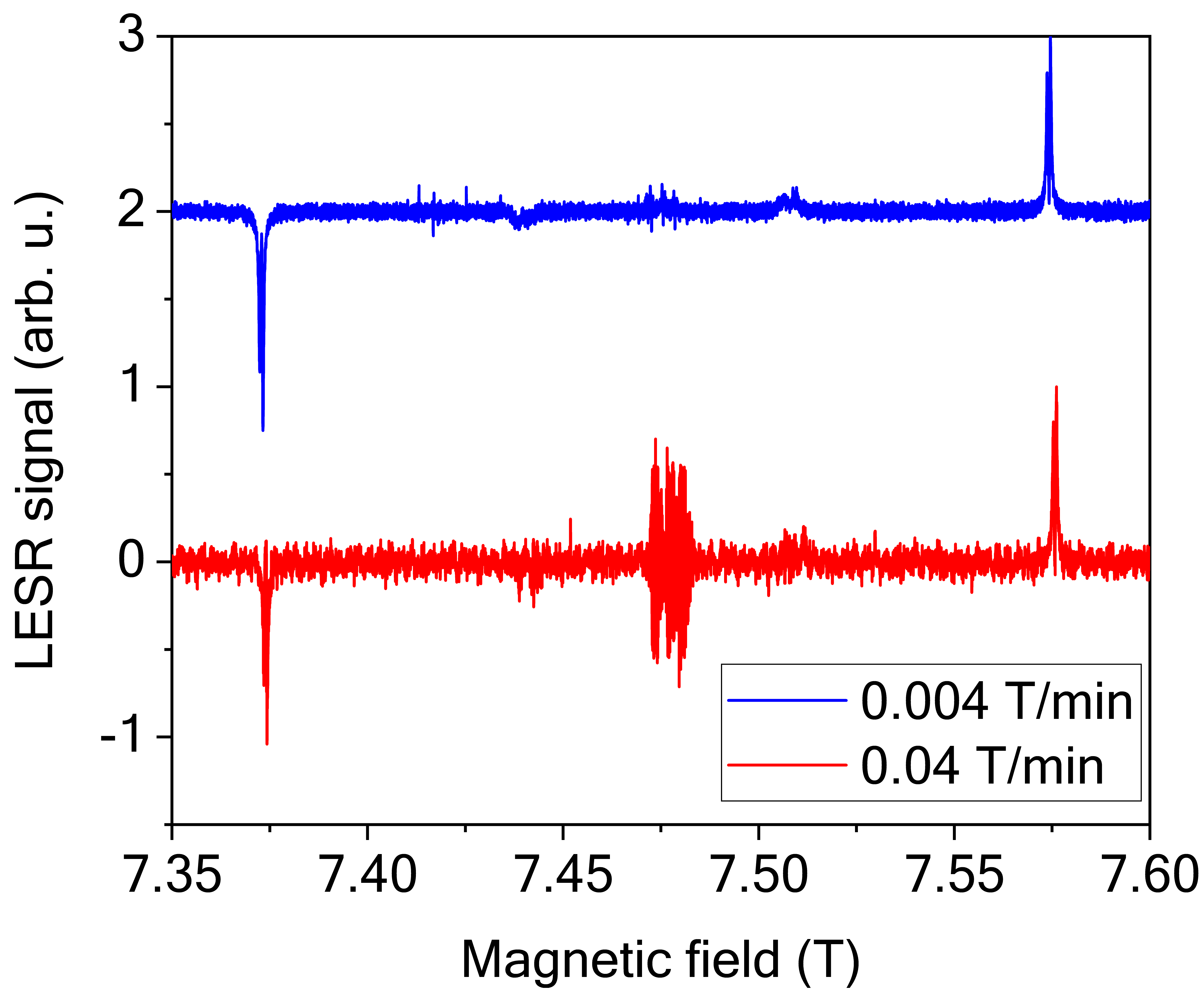}
	\caption{\textbf{Light-induced ESR signal of NV centers}. Both spectra were acquired with $40~\text{Hz}$ optical modulation but with different magnetic field sweep velocities.}
	\label{varying_speed}
\end{figure}

Spectral leakage occurs when $f_\text{chop}$ and $f_\text{sweep}$ are not separated sufficiently. The latter, $f_\text{sweep}$ is obtained from the linewidth, $\Delta B$ (in units of Tesla), and the field sweep speed, $sw$ (in units of Tesla/sec) as ${f_\text{sweep}=sw/\Delta B}$. The typical linewidths of the observed ESR lines are about $0.1~\text{mT}$ and we used field sweeps between ${0.07-0.7~\text{mT/sec}}$, we obtain $f_\text{sweep}=0.7\dots7~\text{Hz}$. This explains why we observe a residual LESR signal for the P1 line when either $f_\text{chop}$ is varied and when it is too small compared to $f_\text{sweep}$ (as shown in Fig. S3), or when $f_\text{sweep}$ is varied and it is too large compared to $f_\text{chop}$ (as seen in Fig. S4).

\begin{figure}[!htb]
	\centering
	\includegraphics*[width=1\linewidth]{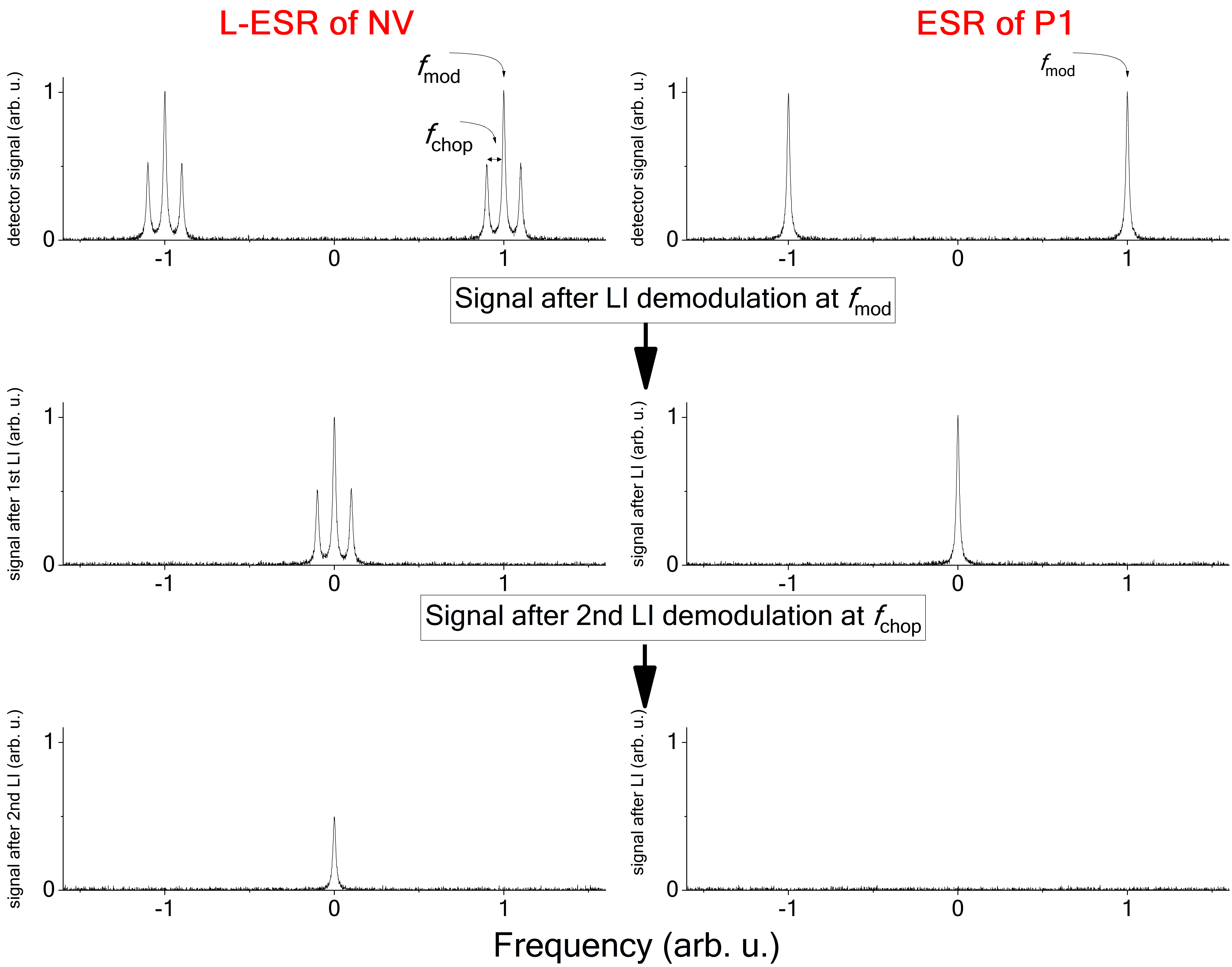}
	\caption{\textbf{Schematical explanation of the observed double lock-in detected signals with varying parameters.}}
	\label{spectral_leakage_explained}
\end{figure}

\section{Angular dependence of the high field ESR of NV centers}

\begin{figure}[!htb]
	\centering
	\includegraphics*[width=1\linewidth]{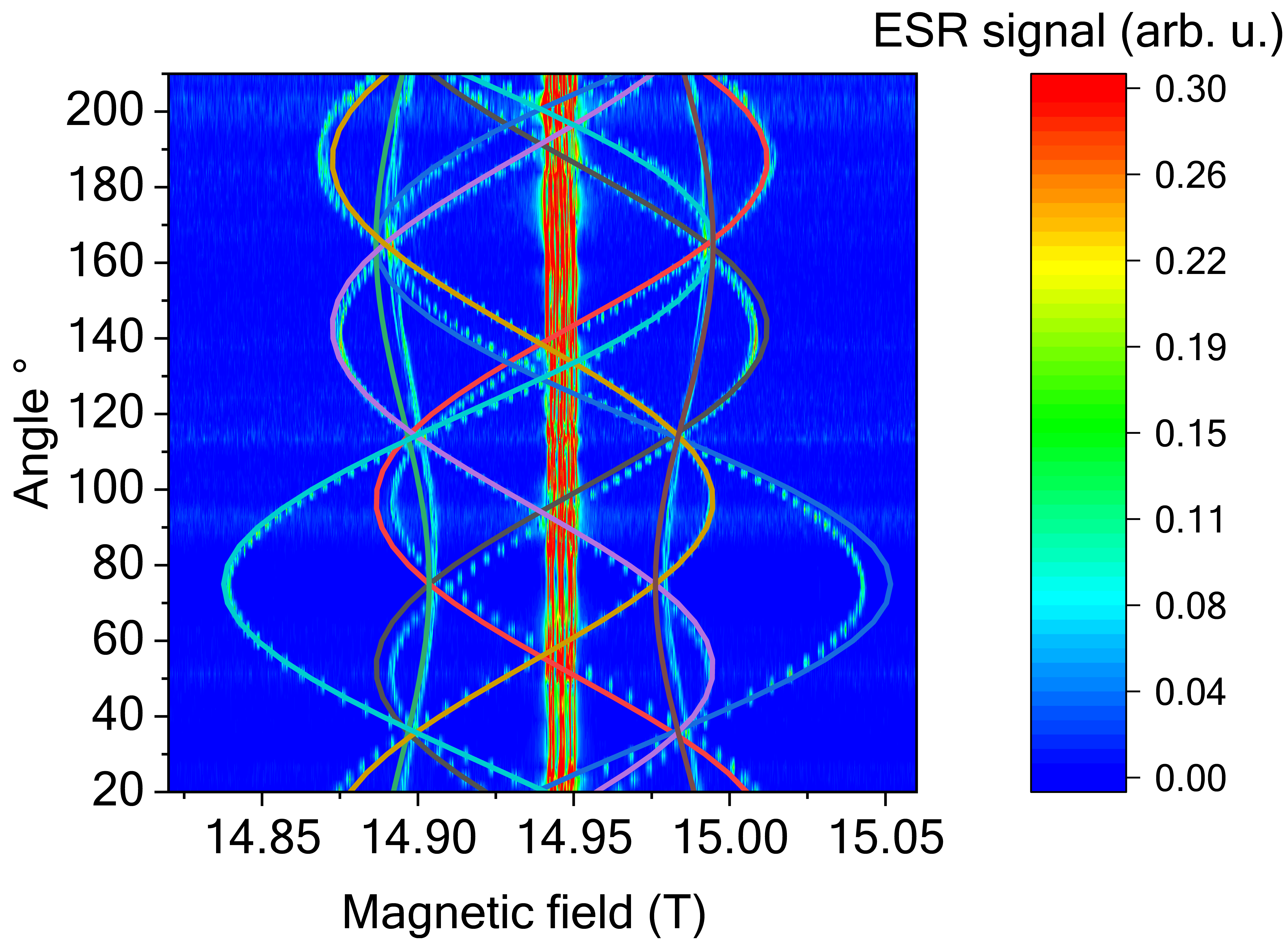}
	\caption{{\color{black}\textbf{Angular dependence of the high field ESR of NV centers.}} When rotated along the $\langle 11\overline{2}\rangle$ crystalline axis the ESR spectrum shows a significant angular dependence. The strong lines around $14.95~\text{T}$ (the $g=2$ resonance position) correspond to the P1 ESR signal. Solid lines are simulated resonance positions.}
	\label{fig:rotation}
\end{figure}

The resonance NV spectra strongly depend on the direction of the magnetic field with respect to the four possible NV-axes. {\color{black}{Figure}} S6 shows the angular dependence for a sample rotated along the $\langle 11\overline{2}\rangle$ axis, compiled from 89 individual ESR spectra.

\section{Visible absorption of the diamond sample used in our studies}

\begin{figure}[!htb]
	\centering
	\includegraphics*[width=1\linewidth]{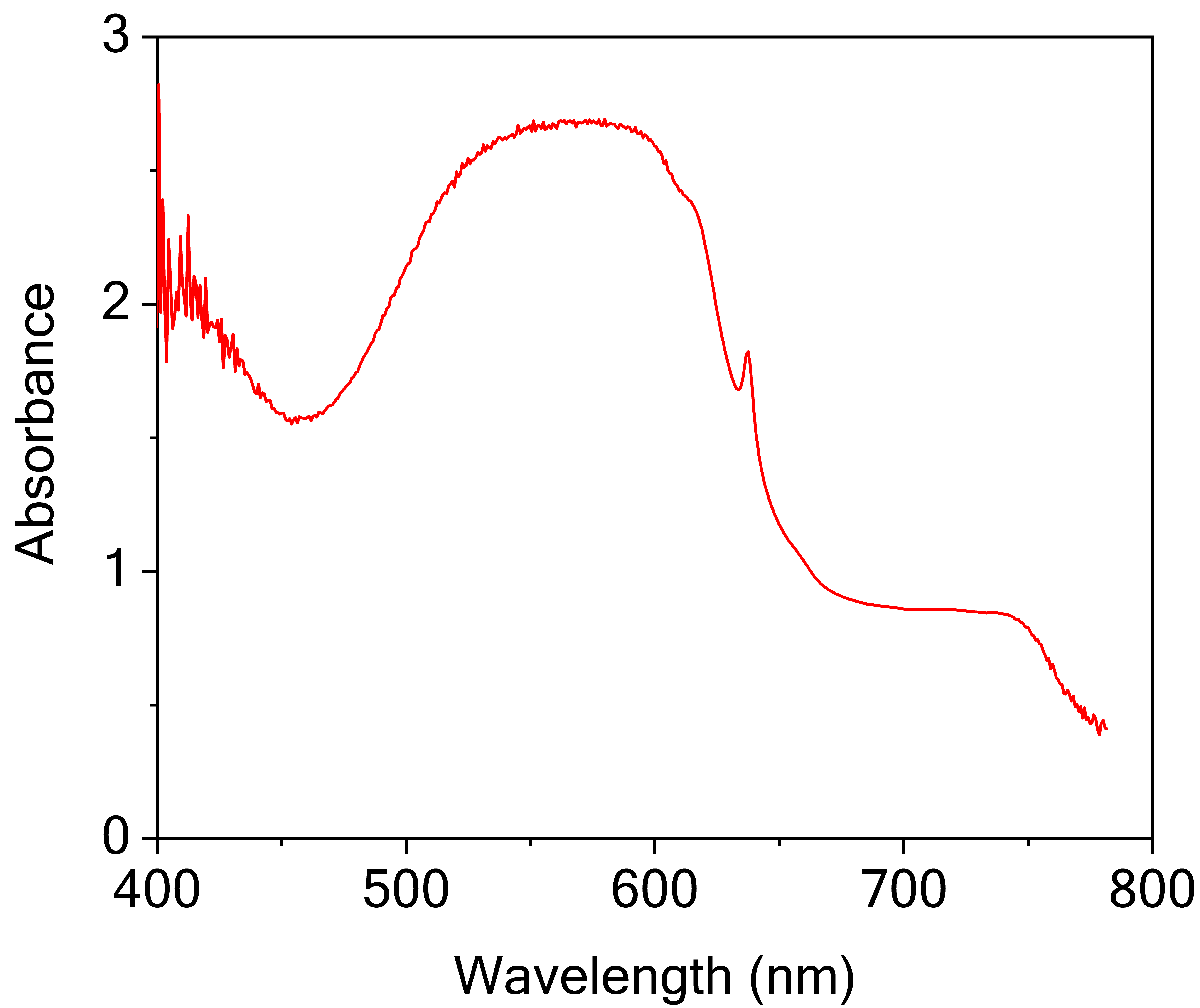}
	\caption{\textbf{Visible absorption spectrum of the diamond sample}. Absorbance of $2.7$ at $532~\text{nm}$ measured on the $0.5~\text{mm}$ thick sample equals an absorption coefficient of $54~\text{cm}^{-1}$.}
	\label{fig:absorption}
\end{figure}

Absorption in the visible range was measured to address the illumination and spin polarization {\color{black}difficulties} of NV centers in the HFESR measurements. {\color{black}{Figure}} S7 shows the absorbance with the characteristic feature of the zero-phonon line of NV centers. 

The absorption is given as the negative logarithm of the transmission, i.e., the ratio of the intensity with ($I$) and without ($I_0$) the sample being in the beam:
\begin{equation}
	A=-\log\left(\frac{I}{I_0}\right).
\end{equation}

The intensity of the light after passing through the sample is:
\begin{equation}\label{eq:expdecay}
	I=I_0 10^{-A}=I_0 10^{-\alpha d}=I_0 \exp(-\sigma n d),
\end{equation}
where $d$ is the thickness of the sample, $\sigma$ is the absorption cross-section and $n$ is the number of NV centers per unit volume. From Eq. \eqref{eq:expdecay} a connection between NV center concentration and the measured absorbance can be given using the absorption cross-section.

The absorbance of our $0.5~\text{mm}$ thick sample at this wavelength is about $2.7$, which yields an absorption coefficient of $54~\text{cm}^{-1}$. {\color{black}The absorption cross-section values of NV centers in the visible regime range between $2.8\times 10^{-17}~\text{cm}^{-2}$ and $9.5 \times 10^{-17}~\text{cm}^{-2}$ reported in Refs. \cite{subedi,savvin,wee2007cross-section,chapman}. The corresponding number of NV centers in a unit volume is between $4.4\times 10^{18}~\text{cm}^{-3}$ and $1.3\times 10^{18}~\text{cm}^{-3}$ (corresponding to $25.2~\text{ppm}$ and $7.4~\text{ppm}$ respectively). We note that the absorption cross-section value of $5.7\pm 0.8\times 10^{-17}~\text{cm}^{-2}$ in reported Ref. \cite{savvin} gives $12.4\ \text{ppm}$ concentration that matches our value of $12\ \text{ppm}$ determined with CW ESR measurement.}

\section{Rotation of ISC transition rates into the laboratory frame in high magnetic field.}

In high magnetic field, the Zeeman interaction is far stronger than that of the ZFS. It is convenient to define the laboratory frame that aligns with the direction of the magnetic field: $\boldsymbol{B}=(0, 0, B)^T$. Zeeman interaction is diagonal thus giving the $\ket{-1}_{\boldsymbol{B}}$, $\ket{0}_{\boldsymbol{B}}$, $\ket{+1}_{\boldsymbol{B}}$ eigenstates separated from each by $g_{\mathrm{e}}\mu_B$  where $\boldsymbol{B}$ in the subscript depicts the magnetic field oriented frame of reference. If the applied $\boldsymbol{B}$-field is oriented towards the $[111]$ symmetry axis then NV's own coordinate system and the laboratory frame would coincide thus the  $\Gamma_{\pm}$, $\Gamma_{z}$ rates shown in Fig. 1 are acting between $\ket{m}_{\boldsymbol{B}}$, where $m={-1,0,+1}$. However, if the $[111]$ oriented $\boldsymbol{B}$-field exerts its effect on one of the three remaining possible NV orientations: $[1\bar{1}\bar{1}]$, $[\bar{1}1\bar{1}]$,  $[\bar{1}\bar{1}1]$ we need to rotate the inter-system crossing (ISC) rates. 

The rotation matrix $R(\alpha,\beta,\gamma)$ can be described by three consecutive rotations depicted by three $(\alpha,\beta,\gamma)$ Euler angles:
\begin{equation}\label{eq:rotation}
	\left(\begin{array}{c}
		0\\
		0\\
		B
	\end{array}\right)_{\boldsymbol{B}}=\underset{R(\alpha,\beta,\gamma)}{\underbrace{R_{\hat{z}}(\alpha)R_{\hat{n}}(\beta)R_{\hat{z}}(\gamma)}}\left(\begin{array}{c}
		B_{x}\\
		B_{y}\\
		B_{z}
	\end{array}\right)_{\text{NV}},
\end{equation}
where we follow the "passive" convention. According to Refs.~\cite{Man_2016,Rose_2018} any irreducible tensor transforming as $S=1$ representation defined within NV's frame can be transformed into the $\boldsymbol{B}$-field's quantization axis by the following Wigner matrix,
%
\begin{equation}\label{eq:WignerMatrix}
	D_{\boldsymbol{B},\text{NV}}^{S=1}(\alpha,\beta,\gamma)=\left(\begin{array}{ccc}
		\frac{1+\cos(\beta)}{2}\expe^{-\cplxi(\alpha+\gamma)} & -\frac{1}{\sqrt{2}}\sin(\beta)\expe^{-\cplxi\alpha} & \frac{1-\cos(\beta)}{2}\expe^{-\cplxi(\alpha-\gamma)}\\
		\frac{1}{\sqrt{2}}\sin(\beta)\expe^{-\cplxi\gamma} & \cos(\beta) & -\frac{1}{\sqrt{2}}\sin(\beta)\expe^{\cplxi\gamma}\\
		\frac{1-\cos(\beta)}{2}\expe^{\cplxi(\alpha-\gamma)} & \frac{1}{\sqrt{2}}\sin(\beta)\expe^{\cplxi\alpha} & \frac{1+\cos(\beta)}{2}\expe^{\cplxi(\alpha+\gamma)}
	\end{array}\right)
\end{equation}
%
that can be used to transform the spin of the three eigenstates as $\ket{m}_{\text{NV}}=\sum_{m^{\prime}}\left[D_{\boldsymbol{B},\text{NV}}^{S=1}(\alpha,\beta,\gamma)\right]_{m,m^{\prime}}\ket{m}^{\prime}_{\boldsymbol{B}}$. Therefore, now we can transform the interaction $\hat{W}$ that drives the non-radiative process between a singlet $\ket{s}$ and the triplet levels $\bra{m}$ with involving the phonon operators $\hat{R}_{m}$ as $\hat{W}= \sum_{m}t^\text{NV}_m\ket{s}\bra{m}_{\text{NV}}\otimes\hat{R}_{m}+ \text{h.c.}$  to create the appropriate amount of phonons that absorb the energy gain by relaxing from the higher electronic state to the lower electronic state and receive the angular momentum difference $m$ between the two electronic states. However, while it seems convenient but surprisingly the aforementioned Wigner transformation on $t_m^{\text{NV}}$ transition amplitudes should be further considered in terms of the phase. The reason is the following. The phase of individual phonons depicted by $\hat{R}_{m}$ is random, thus the amplitudes cannot be added directly. This leads to the random phase approximation,
\begin{equation}\label{eq:RPA}
	\left|t_{m}^{\boldsymbol{B}}\right|^{2}=\sum_{m}\left|D_{\boldsymbol{B},\text{NV}}^{S=1}(\alpha,\beta,\gamma)_{m,m^{\prime}}\right|^{2}\left|t_{m}^{\text{NV}}\right|^{2}\text{,}
\end{equation}
where one has to transform the absolute squares of the transition amplitudes, i.e., the rates with the absolute square of the Wigner matrix as shown below,
\begin{multline*}\label{eq:Wigner}
	\left|D_{\boldsymbol{B},\text{NV}}^{S=1}(\alpha,\beta,\gamma)_{m,m^{\prime}}\right|^{2}=\\
	\left(\begin{array}{ccc}
		\cos^{4}(\beta/2) & \frac{1}{2}\sin^{2}(\beta) & \sin^{4}(\beta/2)\\
		\frac{1}{2}\sin^{2}(\beta) & \cos^{2}(\beta) & \frac{1}{2}\sin^{2}(\beta)\\
		\sin^{4}(\beta/2) & \frac{1}{2}\sin^{2}(\beta) & \cos^{4}(\beta/2)
	\end{array}\right) \text{.}
\end{multline*}
We note that the rotated transition rates are only dependent on the polar angle $\beta$ that rotates the system out from NV's [111] direction and independent on the two azimuthal angles $\alpha$, $\gamma$ that leave the $\boldsymbol{B}$-field's direction intact during their rotation. Therefore, it is not surprising that the transformed rates depend only on the angle between NV's and $\boldsymbol{B}$-field's direction. As mentioned in the main text the external magnetic field is oriented parallel to the symmetry axis of one NV configuration whereas the other three NV configurations exhibit $\beta=109.5^\circ$ angle with respect to the magnetic field direction that leads to the following transformation rules for the intersystem crossing of the three off-axis configurations for the upper branch from $\ket{^3E}$ to $\ket{^1A_1}$,
\begin{equation}\label{eq:Transformedrates_upper}
	\left(\begin{array}{c}
		\Gamma_{\text{ISC}}\\
		0\\
		\Gamma_{\text{ISC}}
	\end{array}\right)_{\text{NV}}\!\!\!\!\!\!\rightarrow\left(\begin{array}{c}
		\Gamma_{+}\\
		\Gamma_{0}\\
		\Gamma_{-}
	\end{array}\right)_{\boldsymbol{B}}=\frac{1}{9}\left(\begin{array}{c}
		5\Gamma_{\text{ISC}}\\
		8\Gamma_{\text{ISC}}\\
		5\Gamma_{\text{ISC}}
	\end{array}\right)_{\boldsymbol{B}}\text{,}
\end{equation}
and for the lower branch acting between $\ket{^1E}$ and $\ket{^3A_2}$,
\begin{equation}\label{eq:Transformedrates_lower}
	\left(\begin{array}{c}
		\Gamma_{\perp}\\
		\Gamma_{z}\\
		\Gamma_{\perp}
	\end{array}\right)_{\text{NV}}\!\!\!\!\!\!\rightarrow\left(\begin{array}{c}
		\Gamma_{+}^{\prime}\\
		\Gamma_{0}^{\prime}\\
		\Gamma_{-}^{\prime}
	\end{array}\right)_{\boldsymbol{B}}=\frac{1}{9}\left(\begin{array}{c}
		5\Gamma_{\perp}+4\Gamma_{z}\\
		8\Gamma_{\perp}+1\Gamma_{z}\\
		5\Gamma_{\perp}+4\Gamma_{z}
	\end{array}\right)_{\boldsymbol{B}}
	\text{.}
\end{equation}
Here, $\Gamma_z\approx 1.2 \times \Gamma_\perp$ according to previous theoretical findings \cite{Thiering2018}. Recently, the ratio between the rates was measured \cite{Reiserer_2016} as  $\Gamma_z\approx 2 \times \Gamma_\perp$. On the other hand, we assume that the transition rates for $\ket{0}$ spin projections are zeroes~\cite{Thiering_2017} for the excited triplet state. The transition rate for $\ket{0}$ is at least a magnitude smaller~\cite{Tetienne_2012, Goldman_2015}  than that of  $\ket{\pm 1}$.  The measured~\cite{Goldman_2015, Goldman_2015_PRB} rate for $\ket{\pm 1}$ is $\Gamma_{\mathrm{ISC}}=2\pi\times \frac{1}{4}(\Gamma_{A_{1}}+2\Gamma_{E_{1,2}})=\frac{2\pi}{4}(16.0+2\times8.3)~\textrm{MHz}=51.2$~MHz while the rate for $\ket{0}$ is $\Gamma_{\mathrm{ISC}}^{(0)}=2\pi\times (0.62\pm0.21)~\textrm{MHz} = 3.90\pm1.32~\textrm{MHz}$. We note that the ratio between $\Gamma_0/ \Gamma_+$ does not change significantly due to this. Eq.~ \eqref{eq:Transformedrates_upper} depicts $\Gamma_0/ \Gamma_+=1.2$ while with additional $\Gamma_{\mathrm{ISC}}^{(0)}$ it would be  $\Gamma_0/ \Gamma_+=1.18$. Therefore, we omit $\Gamma_{\mathrm{ISC}}^{(0)}$ completely from our discussion for simplicity. The result is depicted in Fig. S8.

\begin{figure}[!h]
	\centering
	\includegraphics*[width=0.75\linewidth]{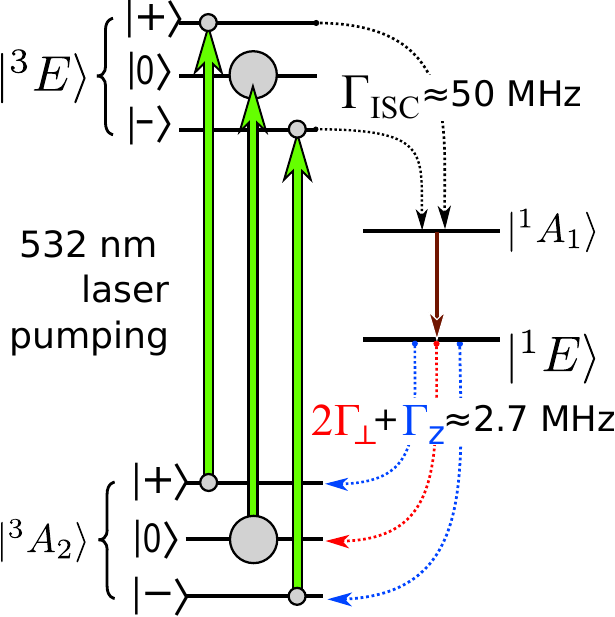}
	\caption{\textbf{Optical spin polarization cycle of NV centers}.}
	\label{fig:isc_Rates}
\end{figure}

Additionally, we assume that the orbital doublets $\ket{^{1}E}$ and $\ket{^{3}E}$ are thermally averaged over their twofold orbital degrees of freedom. Thus, we treat $\ket{^{1}E}$ as a single quantum state effectively from which the $\Gamma_{\perp}$, $\Gamma_{z}$ transition rates originate. Similarly, the upper ISC rate {\color{black}depends} only on the spin wavefunction as the orbital degeneracy in $\ket{^{3}E}$ is averaged out.

As it is described in Eqs.~\eqref{eq:Transformedrates_upper}~and~\eqref{eq:Transformedrates_lower} that the misoriented NV configurations exhibit a much more balanced ISC rate pattern than that of the $\left[111\right]$ oriented NV centers. Therefore, the effect of LESR is not as significant for transitions 2 and 3 (off-axis configurations) as for transitions 1 and 4 (parallel configurations) in Fig. 2b. For the parallel configuration, optical spin-polarization will preferentially populate $\ket{0}$ state which leads to an induced emission for transition "4" ($\ket{0} \rightarrow \ket{-1}$ transition). For the off-axis configurations, the preferential population turns towards $\ket{\pm 1}$ thus induced emission is observed for transition "3" ($\ket{1} \rightarrow \ket{0}$ transition) in Fig. 2b.

\section{\color{black} Outline and critical assessment of a diamond NV based TASER system.}

{\color{black}
	The construction of a THz laser (TASER in the following) system based on diamond NV centers could closely follow a microcavity laser design. The proposed design is shown in Fig. S9.
	
	\begin{figure}[!h]
		\centering
		\includegraphics*[width=1\linewidth]{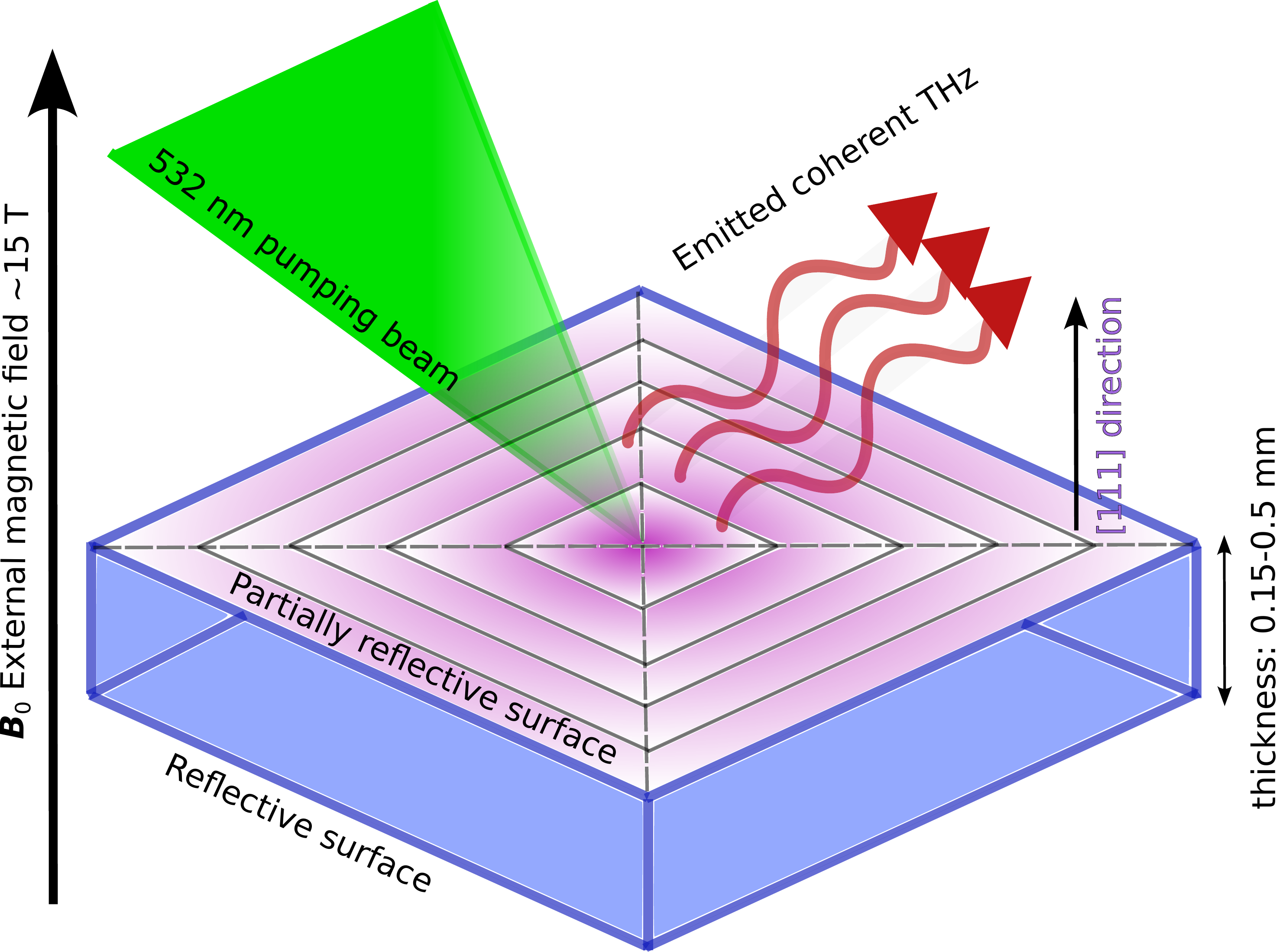}
		\caption{\textbf{Proposed construction of a $\boldsymbol{0.4}\ \text{THz}$ TASER.}}
		\label{fig:TASER_design}
	\end{figure}
	
	The magnetic resonance selection rules dictate that the polarization plane of the absorbed or generated THz radiation is perpendicular to the external magnetic field. Efficient lasing was observed with such a geometry in Ref. \cite{Breeze2018}. This is the same geometry in which we observed the THz emission in the diamond NV centers. 
	
	More specifically, this is a so-called Fabry-Pérot microcavity design \cite{microcavity}. The working condition for the Fabry-Pérot microcavity is $\lambda=2 L/m$, where $L$ is the thickness of the microcavity, $m$ is a positive integer, $\lambda$ is the wavelength of the generated radiation in the medium.
	
	The proposed geometry is also optimal for the pumping efficiency, as the polarization of the incoming (pumping) light is perpendicular to the NV axis \cite{alegre2007polarization}. We focus on a TASER operating at $0.4\ \text{THz}$, as it can be achieved with a conventional laboratory magnetic field of around $15\ \text{T}$. The corresponding wavelength is $\lambda(\text{0.4\,THz})=0.314\,\text{mm}$ using the index of refraction $n=2.38$ in the THz range for diamond \cite{rogalin2018optical}. This means that a microcavity, which sustains 0.4 THz TASER modes, has a thickness of $L=0.157\,\text{mm}$ or its multiples. 
	
	In fact, a thickness between 0.15-0.5 mm is ideal for a sample with the NV concentration such as in our sample due to the above described $54\ \text{cm}^{-1}$ absorption coefficient measured at $532\ \text{nm}$. This means that a sample with a thickness of 0.2 mm absorbs $1/e$-th part of the exciting laser. The pumping light would not be exploited optimally for a thinner sample. For a much thicker sample, the NV centers away from the sample surface would be pumped less efficiently by the attenuated power of the pump laser.
	
	The key factor in determining whether the TASER operation can be sustained is the cooperativity \cite{Breeze2018} factor: $C=2 P N \cdot T_2^*$. Here, $P$ is the Purcell factor, $N$ is the number of active NV centers and $T_2^*$ is the spin dephasing factor. The latter usually describes the inverse of the observable linewidth in magnetic resonance experiments. The Purcell factor is well approximated by
	$P\approx 0.1 Q \frac{\lambda^3}{V}$, where $Q$ is the quality factor of the microcavity resonator, $V$ is its volume and $\lambda$ is the wavelength of the emitted radiation inside the cavity.
	
	While obtaining absolute estimates for these values is difficult, we make a comparison with Ref. \cite{Breeze2018}, where microwave lasing (or masing) was successfully achieved. The concentration in our sample is about 30 times larger than in Ref. \cite{Breeze2018} we thus estimate that the number of active NV centers is similarly larger herein. We estimate that the Purcell factor for the proposed TASER design should be similar to that in Ref \cite{Breeze2018}. It goes essentially with the volume filling factor of the sample with respect to the resonator, which sustains the radiation. For both the maser and TASER designs, the two are around unity. Typical microwave quality factors are a few thousand which is also feasible in microcavity resonators. Therefore, we expect a similar Purcell factor for the two designs. The last remaining factor, which requires careful consideration is the dephasing rate. In electron spin resonance on solids, the dephasing rate or $1/T_2^*$ arises mainly from impurities and unresolved hyperfine couplings \cite{Portis}. This is markedly different from the case of nuclear magnetic resonance (especially for liquids) where the dominant contribution to $1/T_2^*$ is due to the magnetic field inhomogeneity. Were this also the case herein, one would expect that the higher magnetic field results in a much higher dephasing rate (about 16 T/0.3 T$\approx$ 50 times higher). However, the dephasing effects of impurities or unresolved hyperfine couplings are inherently field \emph{independent} \cite{AbragamBook}, we therefore expect similar dephasing rates for the proposed TASER than in Ref. \cite{Breeze2018}. Altogether, this consideration gives that about 30 times higher cooperativity could be achieved in our proposed design, which predicts the operation of an efficient THz laser.
}

\end{document}